\documentclass[useAMS,usenatbib]{mn2e}
\usepackage{graphicx}
\usepackage{subfigure}
\usepackage{amsmath}
\usepackage{epsfig}
\usepackage{epstopdf}
\usepackage{amssymb}

\title[Mass-to-light ratio of UGC11919]{Long-slit spectral observations and
stellar mass-to-light ratio of spiral galaxy UGC11919}

\author[A. Saburova, A. Zasov, R. Uklein, I. Katkov]
{A. Saburova$^{1}$\thanks{Corresponding author\,: {\tt saburovaann@gmail.com}}, 
A. Zasov$^{1,2}$, 
R. Uklein$^{3}$,
I. Katkov$^{1}$\\
$^1$ Sternberg Astronomical Institute, 
Moscow M.V. Lomonosov State University, 
Universitetskij pr., 13,  Moscow, 119992, Russia \\
$^2$ Department of Physics, 
Moscow M.V. Lomonosov State University, 
Leninskie Gory, Moscow, 119991, Russia\\
$^3$ Special Astrophysical Observatory, 
Russian Academy of Sciences, 
Nizhniy Arkhyz, Karachai-Cherkessian Republic, 357147, Russia \\
}

\begin{document}

\pagerange{\pageref{firstpage}--\pageref{lastpage}} \pubyear{2002}

\maketitle

\label{firstpage}

\begin{abstract}

We performed the long-slit observations of spiral galaxy UGC11919 with the
Russian 6-m telescope to study its kinematics and stellar population. The
previous studies gave basis to suspect that this galaxy possesses a peculiarly
low mass-to-light ratio $M/L_B$ of stellar population which could indicate the presence of bottom-light stellar initial mass function (IMF). The ratio $M/L_B$ estimated for different
evolutionary models of stellar population using both the broad-band magnitudes and the detailed
spectral data confirms this peculiarity if the disc inclination angle $i \ge
30\degr$, as it was obtained earlier from the optical photometry, in a good agreement with the
H{\sc i} data cube modelling. However the re-processing of H{\sc i} data cube we carried out showed that it is compatible with
much lower value  $i\approx 13 \degr$ corresponding to the "normal" ratio $M/L_B$, which does not need any peculiar stellar IMF. Stellar velocity dispersion measured at one disc radial scalelength from the center also better agrees with the low disc inclination. However in this case we should admit that the disc possesses a non-axisymmetric shape even after taking into account a two-armed spiral structure.

The derived stellar kinematic profiles reveal a signature of
kinematically decoupled nuclear disc in the galaxy. Using different evolution models of stellar population we estimated the stellar metallicity [Z/H] (-0.4, -0.5  and -0.3 dex) and the  mean
luminosity-weighted  (for the luminosity in the spectral range $4800-5570$ \AA)  stellar age  (4.2, 2.6 and 2.3 Gyr) for the bulge, disc and nuclear disc of this
galaxy respectively.   

\end{abstract}

\section{Introduction}

One of the key problem of formation and evolution of galaxies 
is the stellar initial mass function (IMF). At the present
time the question of the universality of  IMF  still remains open (see f.e. the discussion in  \citealt { Kroupa2002, Gilmore, Bastian, Gunawardhana}).


It cannot be excluded that  in some discy galaxies IMF may differ significantly from that usually accepted as "typical" or "normal" IMF observed in our Galaxy or in the neighbor galaxies. In the current article, we define as the "normal" IMF  the scaled Salpeter (\citealt{imf}, \citealt{bdj}) or Kroupa (\citealt{imf2}) IMFs which we used in the spectral fitting. 

In \cite{Saburova2009} we attempted to identify galaxies with presumably  non-typical IMF by choosing the objects where the rough estimate of the dynamical mass-to-light ratios $M/L_B= v^2r_{25}/GL_B$, where $v$ is the velocity of rotation and  $r_{25}$ is the optical radius, strongly disagrees with $M/L_B$ ratios of stellar populations found from the color indices. The most interesting cases are where galaxies possess a dynamical mass that is too low for their luminosity and color, because the presence of a dark halo in these galaxies may only increase the discrepancy between the dynamic and photometric mass estimates. In many cases this discrepancy is the result of errors,  connected mostly with the estimation of dynamical masses, hence every case should be verified using the additional data. It is not surprising that the analysis of measurement errors has significantly reduced the final list of potential candidates for galaxies with peculiar mass-to-luminosity ratios.
In  \cite{saburovaetal2013} (hereafter Paper I) we performed a detailed study of two objects from this list. It was concluded that for one of them -- UGC11919 -- the peculiarly low dynamical mass-to-light ratio concords with new optical and H{\sc i} observations. 

UGC11919 is SABb galaxy with
apparent magnitude corrected for Galactic extinction $m_B=13.5$ mag. Its
cosmological velocity (5576 {$\textrm{km~s$^{-1}$}$) corresponds to the distance 74 Mpc and
luminosity $M_B = -20.8$ mag. The galaxy possesses the bar, two prominent symmetric
spiral arms, and, as our  H{\sc i} observations   showed, a giant hydrogen
ring of low intensity, surrounding the optical galaxy. According to Paper I, the best fit
dynamical mass-to-light ratios $M/L_B$ of stellar disc and bulge
($(M/L_B)_{disc}=0.5$, $(M/L_B)_{bulge}=0.4$) are correspondingly 3 and 5
times  lower, than it is expected from the model $M/L_B$-color
relations  for the scaled Salpeter IMF proposed by
\cite{bdj}. 

However, the inclination angle estimate  is quite uncertain for this galaxy. It leaves the question of mass estimates of UGC11919 open. Indeed, Hyperleda database gives the inclination angle $i=57\degr$ obtained from the outer isophotes axes ratio. More elaborate 2D-photomerty carried out in Paper I gave i = $58\degr\pm 4\degr$, leading to the peculiarly low $M/L_B$.  After masking the
prominent spiral arms the flattening of optical
isophotes ($i=\arccos(a/b)$) for $r\le 57 \arcsec$ 
admits the inclination $i \ge 35 \degr$. 
Thus the photometric estimations of $i$ do not solve the problem of low dynamical mass of stellar population. 

The H{\sc i} data analysis partially softened the discrepancy, giving the lower value $i\approx 30\degr$ for the model taking into account the velocity and density perturbations due to two-armed spiral structure. In Paper I we also considered the warped
disc model where the inclination and the position angle varied with radius,
representing the symmetric S-shaped warp. However in this model the
inclination changes only moderately remaining between 45 and 35 degrees. So both models give the range of $i$ that does not eliminate the disagreement with the $M/L_B$ ratio expected from the color indices of the galaxy. 

To reconcile the
dynamic and photometric mass estimates the
inclination $i$ should be about $14\pm 1.3\degr$. As we show below this value is not in conflict with the observational data. 

In this paper we re-processed the H{\sc i}  data cube of UGC11919 and attracted the data of spectral observations we carried out at 6m  telescope BTA to clarify the dynamic properties of this galaxy and to study the properties of its stellar population.

\section{The inclination angle: re-processing of H{\sc i} data }

\begin{figure*}
\centering
\includegraphics[width=0.4\textwidth]{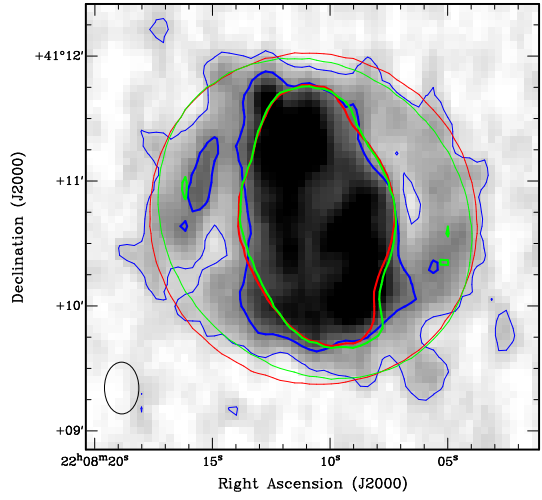}
\caption{ H{\sc i} total intensity map with   the overplotted observed column 
density 5.4, 2.0 $\times 10^{19}$  ${\rm atoms \, cm}^{-2}$ contours (blue), 
in comparison with the model contours for the same column densities, 
reproduced in the  bisymmetric models for $i=30\degr$ (green) and $i=13\degr$ (red).} 
\label{twomodels}
\end{figure*}
\begin{figure*}
\centering
\includegraphics[width=0.4\textwidth]{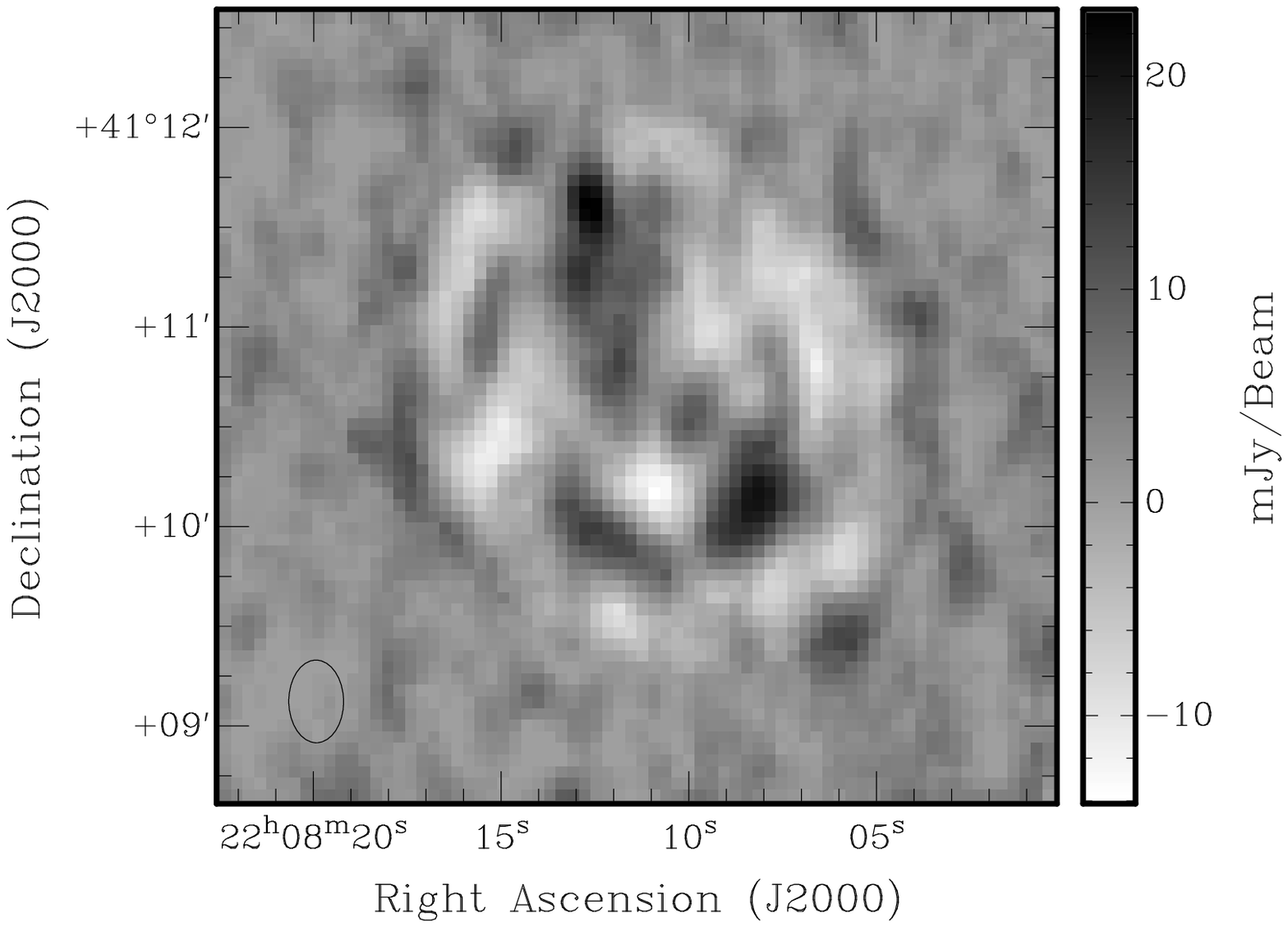}
\includegraphics[width=0.4\textwidth]{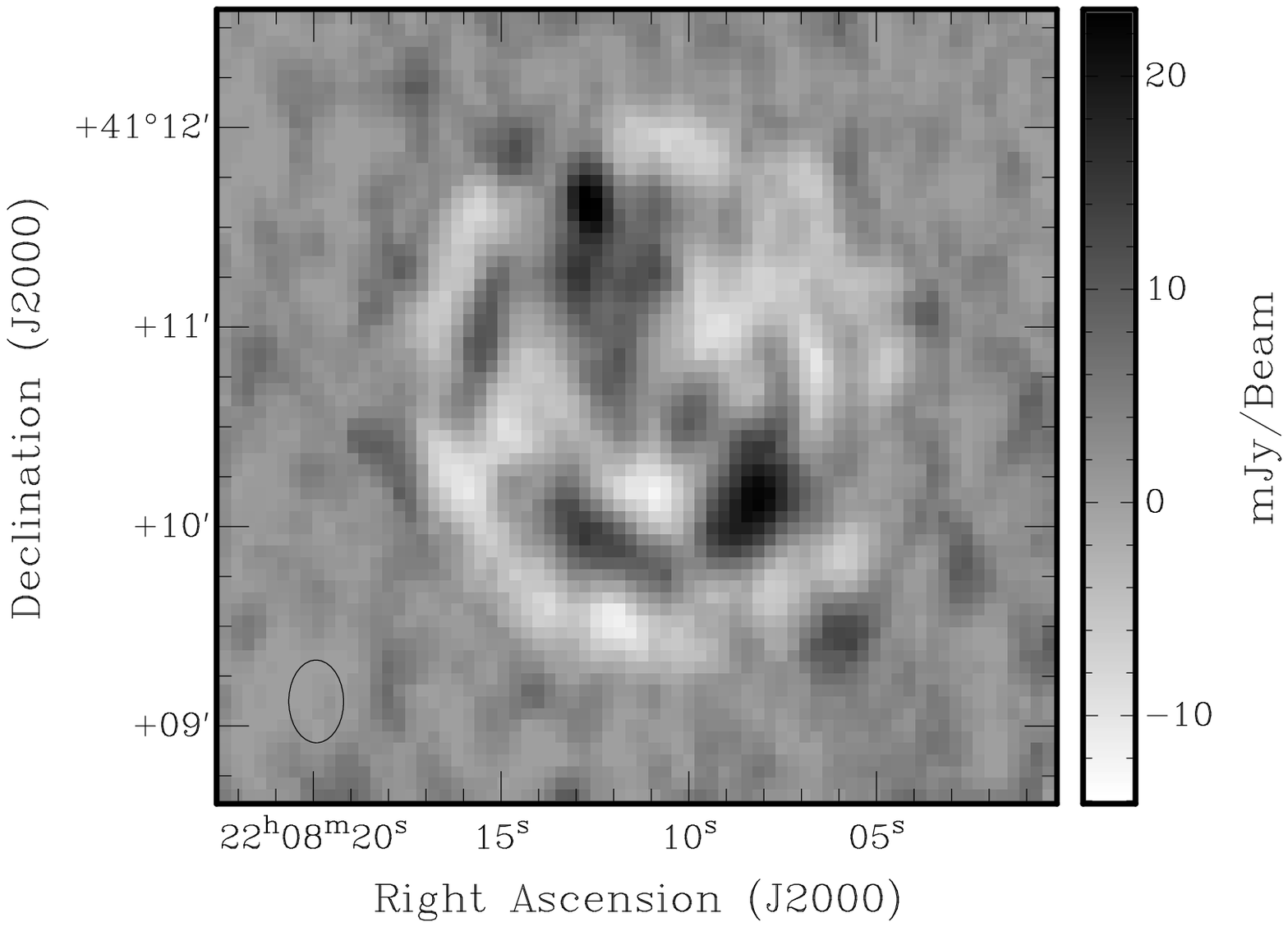}
\caption{ H{\sc i} total intensity map residuals (observed map minus model map) for the  bisymmetric models for $i=30\degr$ (left) and $i=13\degr$ (right). } 
\label{m0_residuals}
\end{figure*}
\begin{figure*}
\centering
\includegraphics[trim= 0 120 0 70,clip,width=0.5\textwidth]{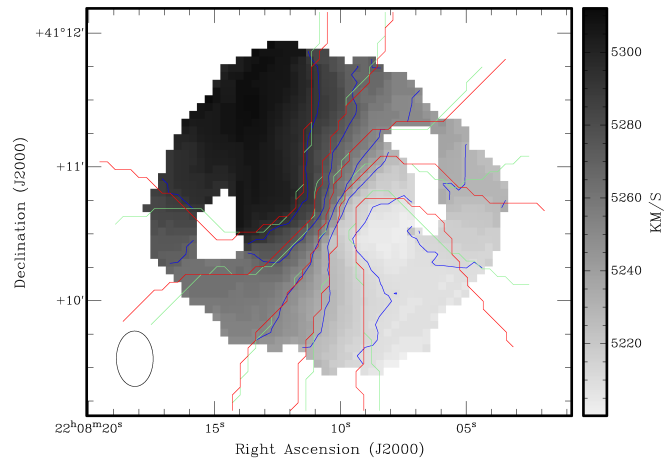}

\caption{ The observed first-moment map of UGC11919 with the overplotted observed (blue) and 
bisymmetric model (green for $i=30\degr$ and red  for $i=13\degr$) contours 
$v=v_{sys}\pm$ 0, 20, 40 $~  \textrm{km~s}^{-1}$.} 

\label{m1comparison}
\end{figure*}
\begin{figure*}
\centering
\includegraphics[trim= 0 120 0 70,clip,width=0.4\textwidth]{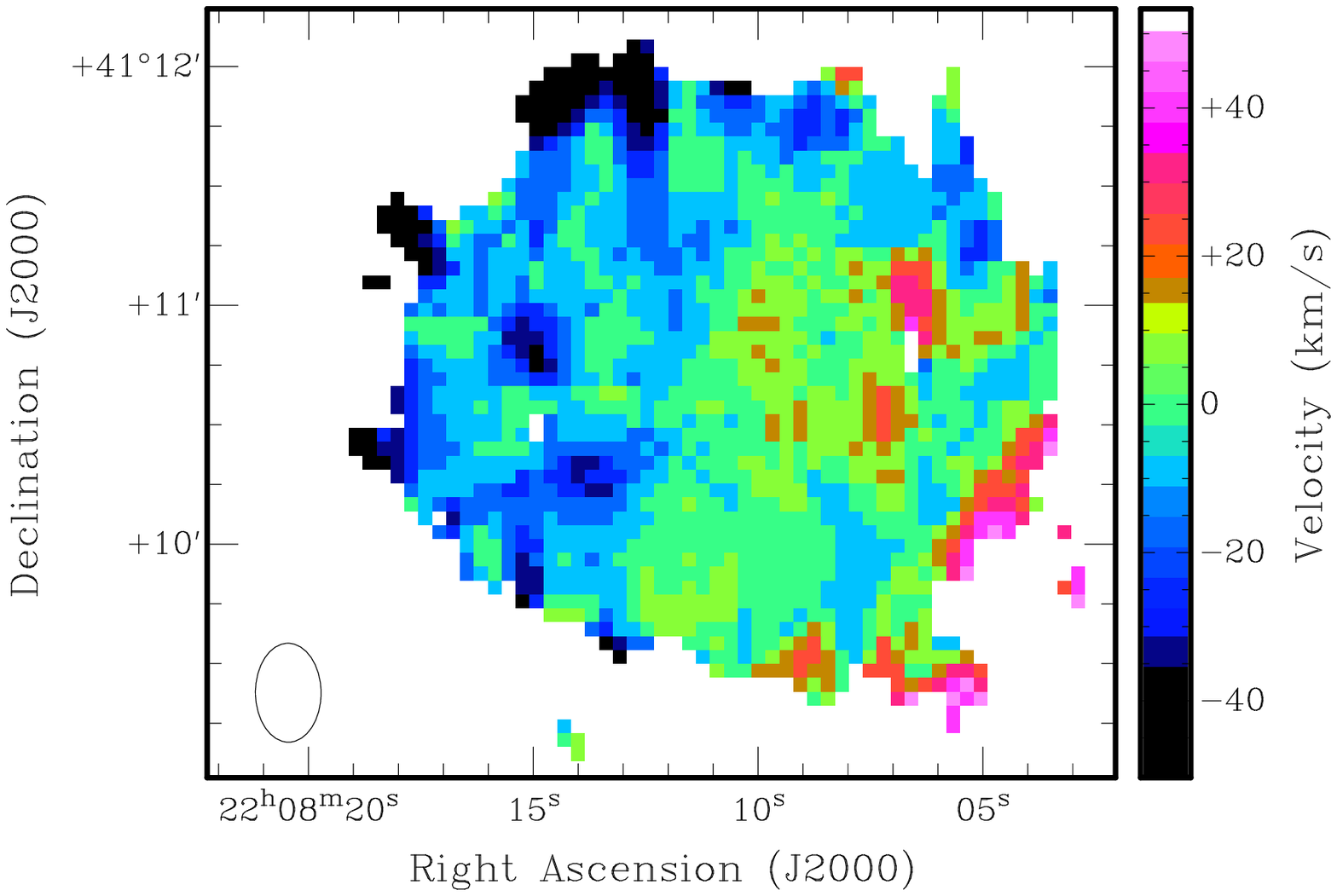}
\includegraphics[trim= 0 120 0 70,clip,width=0.4\textwidth]{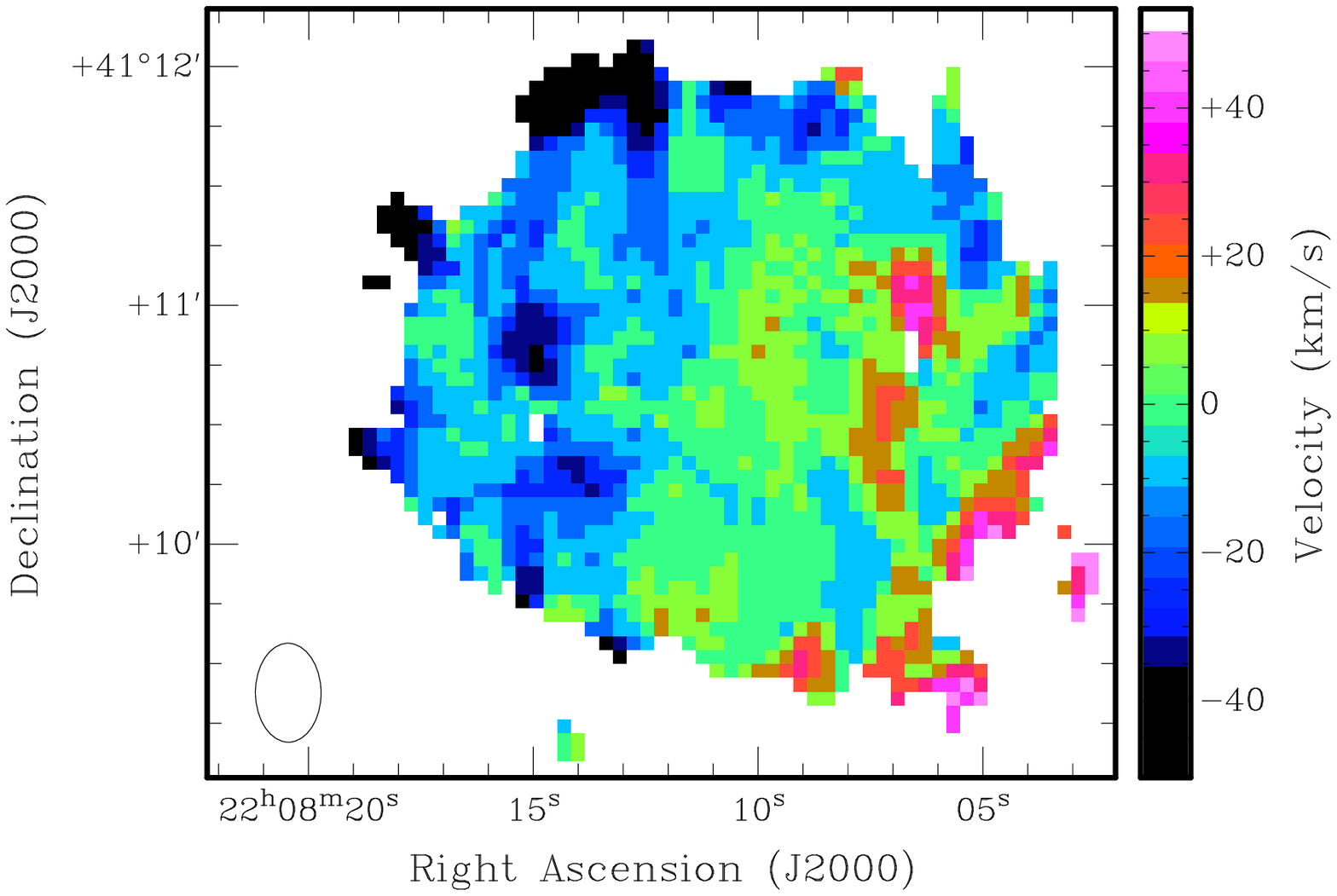}
\caption{ The residuals of the first-moment map of UGC11919 (observed minus model maps) for the
bisymmetric model with $i=30\degr$ (left) and  $i=13\degr$ (right). } 
\label{m1_residuals}
\end{figure*}

Observations of UGC11919 in the H{\sc i} line mentioned above were carried out earlier 
in the Westerbork Synthesis Radio Telescope (WSRT). The data processing and the results were described earlier
in Paper I. In the current work we applied the same program
TiRiFiC\footnote{http://www.astron.nl/~jozsa/tirific/} to reprocess the H{\sc i} data
cube obtained in order to check the compatibility of kinematical
model with low inclination angle. TiRiFiC
is a software allowing for a direct fit of modified tilted-ring models to data
cubes \citep{J2007}. We used the model similar to the bisymmetrical model
described in  Paper I for a thin disc, which takes into account the presence
of the ordered non-circular motions in addition to morphological deviations from axial
symmetry from two arm spiral structure. This time  we changed the initial guesses of the parameters significantly in order to get as low inclination as possible which remains compatible with the H{\sc i} data cube and obtained a model with the inclination $i=13\degr$ in a good agreement with the value needed to eliminate the discrepancy between the dynamical and photometrical mass estimates. 

In Figs. \ref{twomodels}, \ref{m0_residuals} we show the observed H{\sc i}
total intensity map with the overplotted observed isodensities, the model
contours and the residuals (observed minus model maps) for high and low inclination ($30 \degr$ and $13 \degr$). 
In Fig. \ref{m1comparison} we also compare the observed and model first moment H{\sc
i} maps. The contours in Fig. \ref{m1comparison} correspond to the velocities
$v=v_{sys} \pm$ 0, 20, 40 $ \textrm{km~s}^{-1}$ of the observed (blue) and model maps for
$i=13\degr$ (red) and $i=30\degr$ (green). In Fig. \ref{m1_residuals} we demonstrate the residues of the models from the observed first moment map.  One may see that model with low inclination slightly overpredicts the rotational amplitude and the surface density of the gas in the peripheral regions, although it can't be rejected. 
A similar conclusion may be done from the comparison of the model and the observed data cubes
(see Fig. \ref{datacube}). Hence the H{\sc i} data do not contradict to the low inclination and to the non-peculiar IMF.


We checked the compatibility of the given inclination angles with the position of the galaxy in the Tully-Fisher diagrams. In Fig.\ref{tf} we plotted  UGC11919 assuming two values of inclination ($i=30\degr$ and $i=13\degr$) on the three kinds of Tully-Fisher (TF) diagrams: a classical one, where the rotational velocity is compared to the B-band luminosity (Fig.\ref{tf}, left), baryonic (Fig.\ref{tf}, centre) and stellar  (Fig.\ref{tf}, right) masses of galaxies. The parallel lines  in these figures mark the sequences found by \citet{McGaugh2005} and \citet{McGaugh2015} for the samples of disc galaxies and their uncertainties. The luminosity and masses of UGC11919 are taken from Paper I and Sect. \ref{mm} of current paper (for $i=13\degr$). As could be seen from Fig.\ref{tf} the low inclination case nicely agrees with the relations followed by spiral galaxies.
 
 We came to conclusion that the disc of UGC11919 either is really too light for its luminosity, or the galaxy is seen nearly face-on. In the latter case the isophotes flatness reflects a large-scale non-axisymmetry of the disc. The oblate form of the optical disc (the internal axial ratio of $(b/a)_0=0.89$ or lower) is needed to explain the difference between the optically observed inclination $i\ge 35 \degr$ and the inclination  $i=13 \degr$ corresponding to the "normal" $M/L_B$ ratio. In both cases the galaxy appears to be interesting for further studying. The aim of current article is to test the results of dynamical modeling obtained in Paper I and to explore the properties of stellar population of the galaxy using the long-slit data.

 In current paper we give more preferences to the inclination $i=13 \degr$ obtained from the H{\sc i} data cube modeling and corresponding to the "normal" IMF since it is in good agreement with the stellar velocity dispersion data (see below). However one should
keep in mind that the higher inclination cannot be fully excluded.  

%
%

\begin{figure*}
\centering
\includegraphics[width=0.3\textwidth,keepaspectratio]{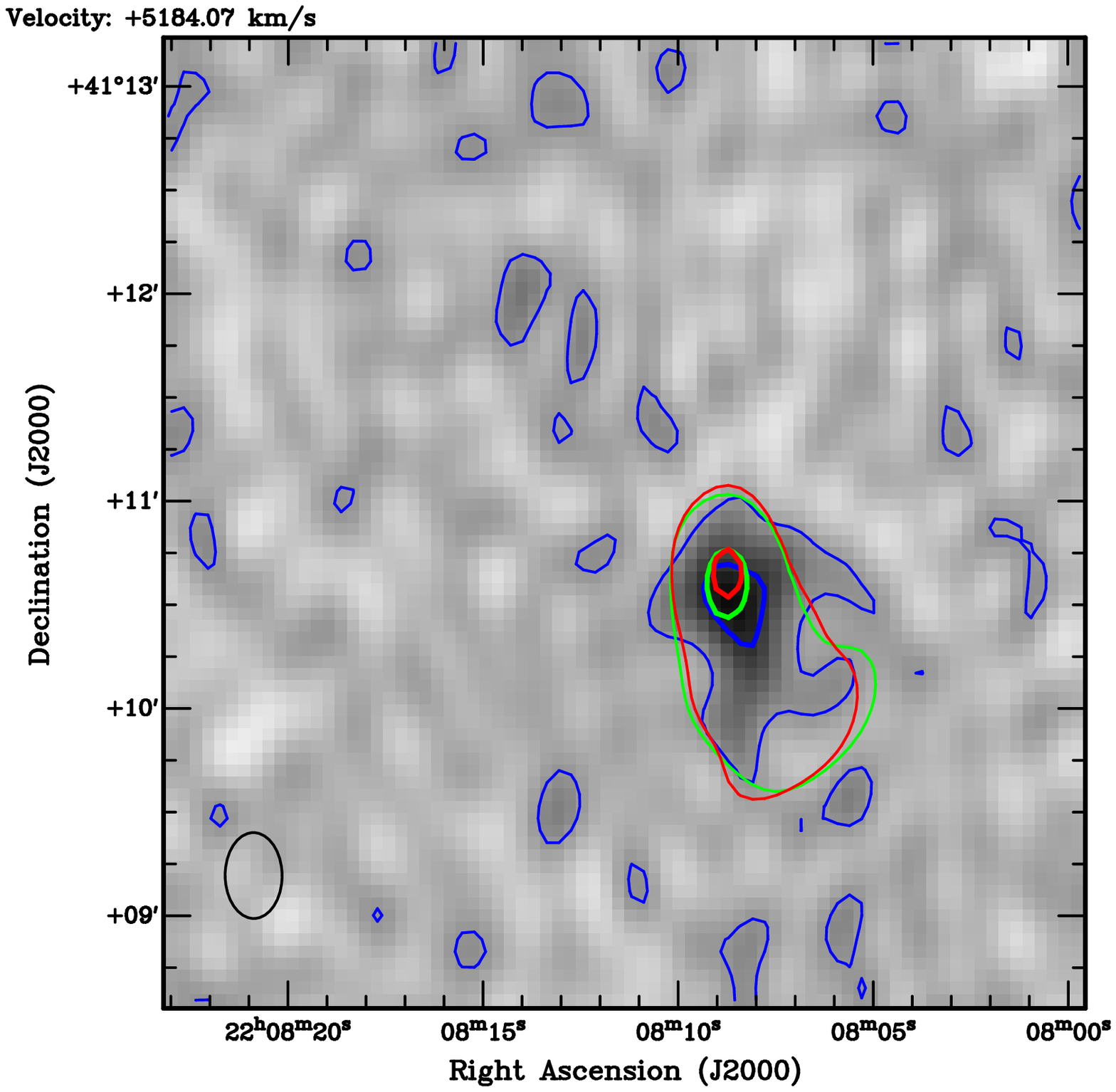}
\includegraphics[width=0.3\textwidth,keepaspectratio]{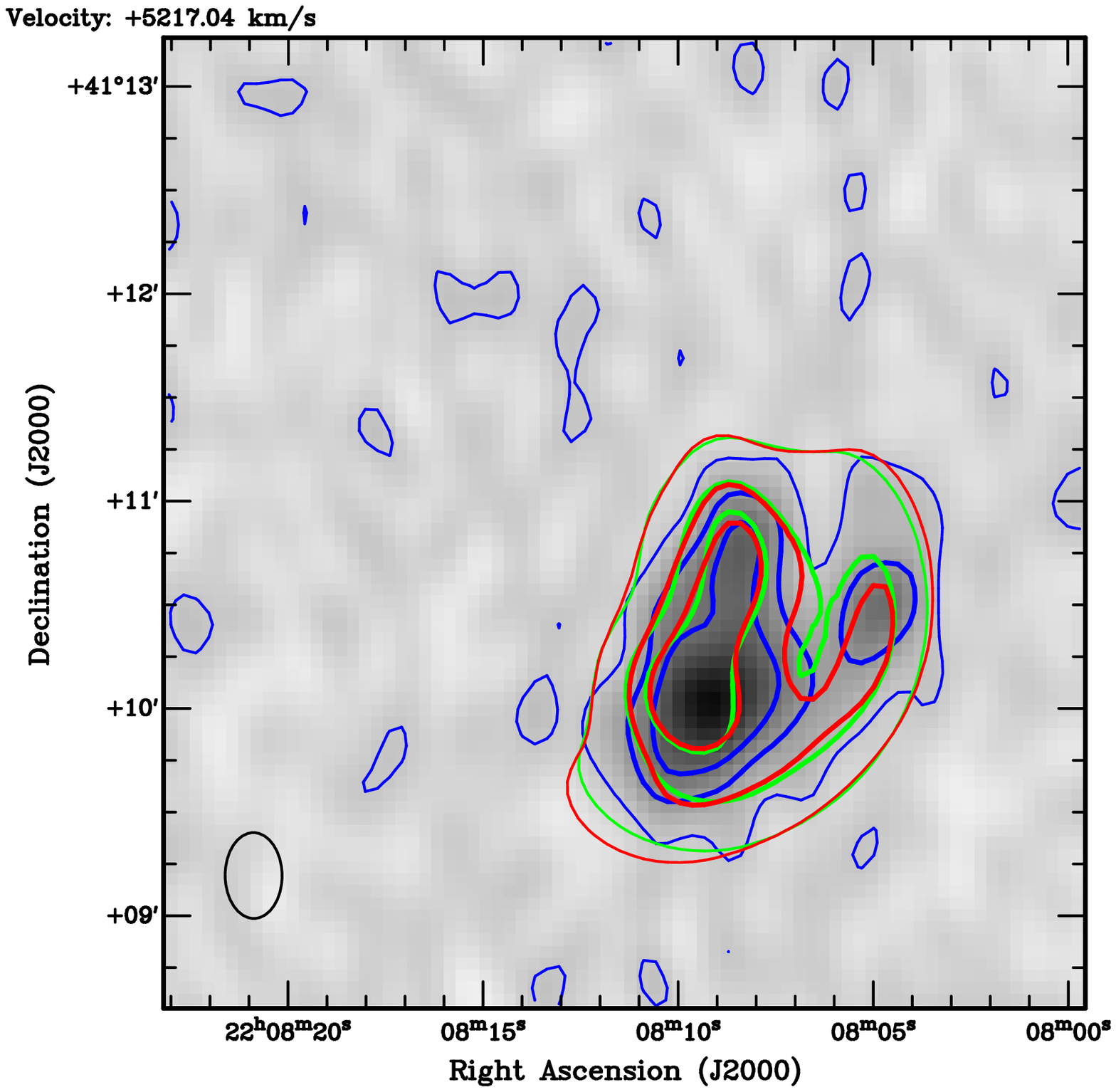}
\includegraphics[width=0.3\textwidth,keepaspectratio]{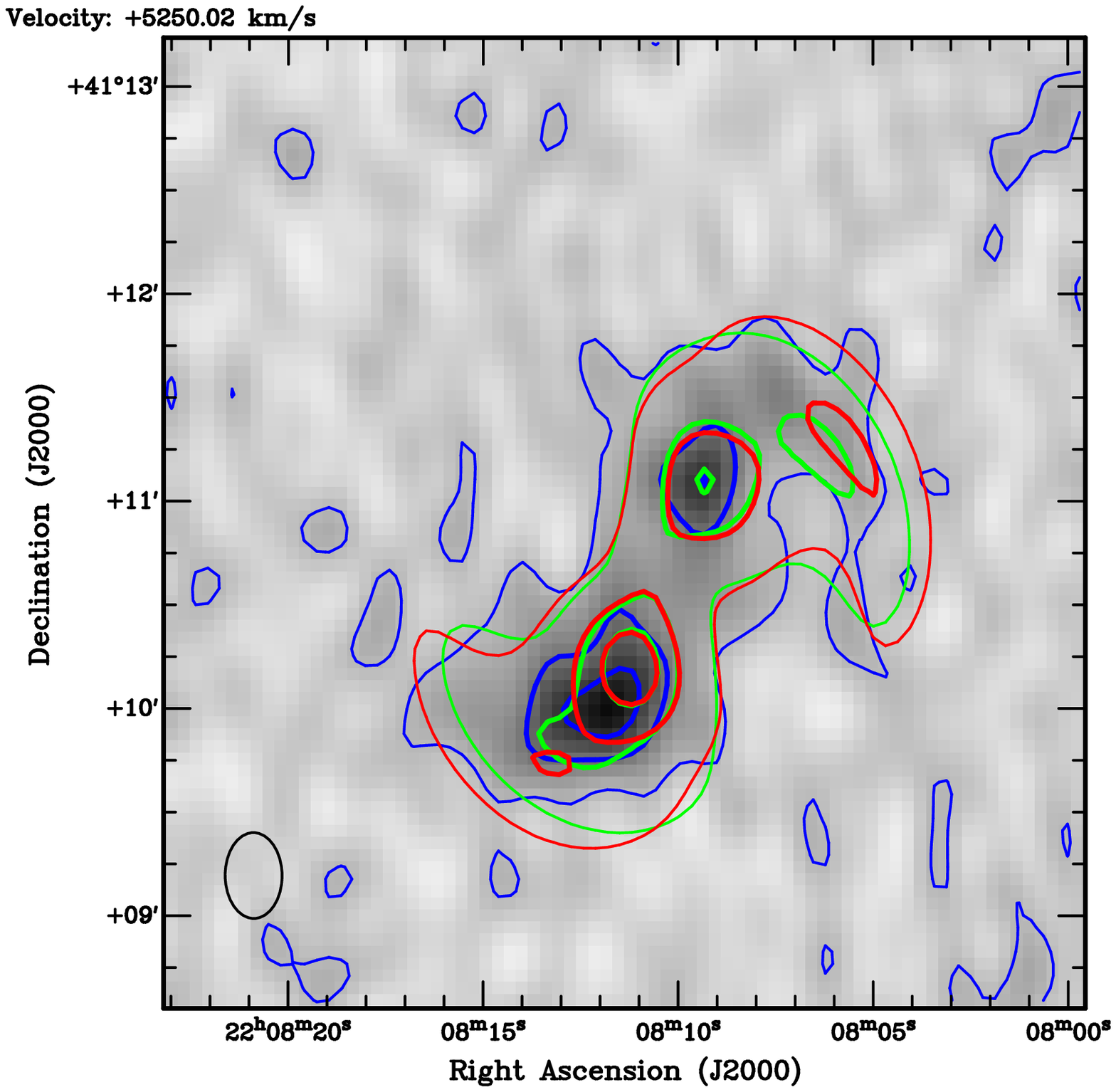}
\includegraphics[width=0.3\textwidth,keepaspectratio]{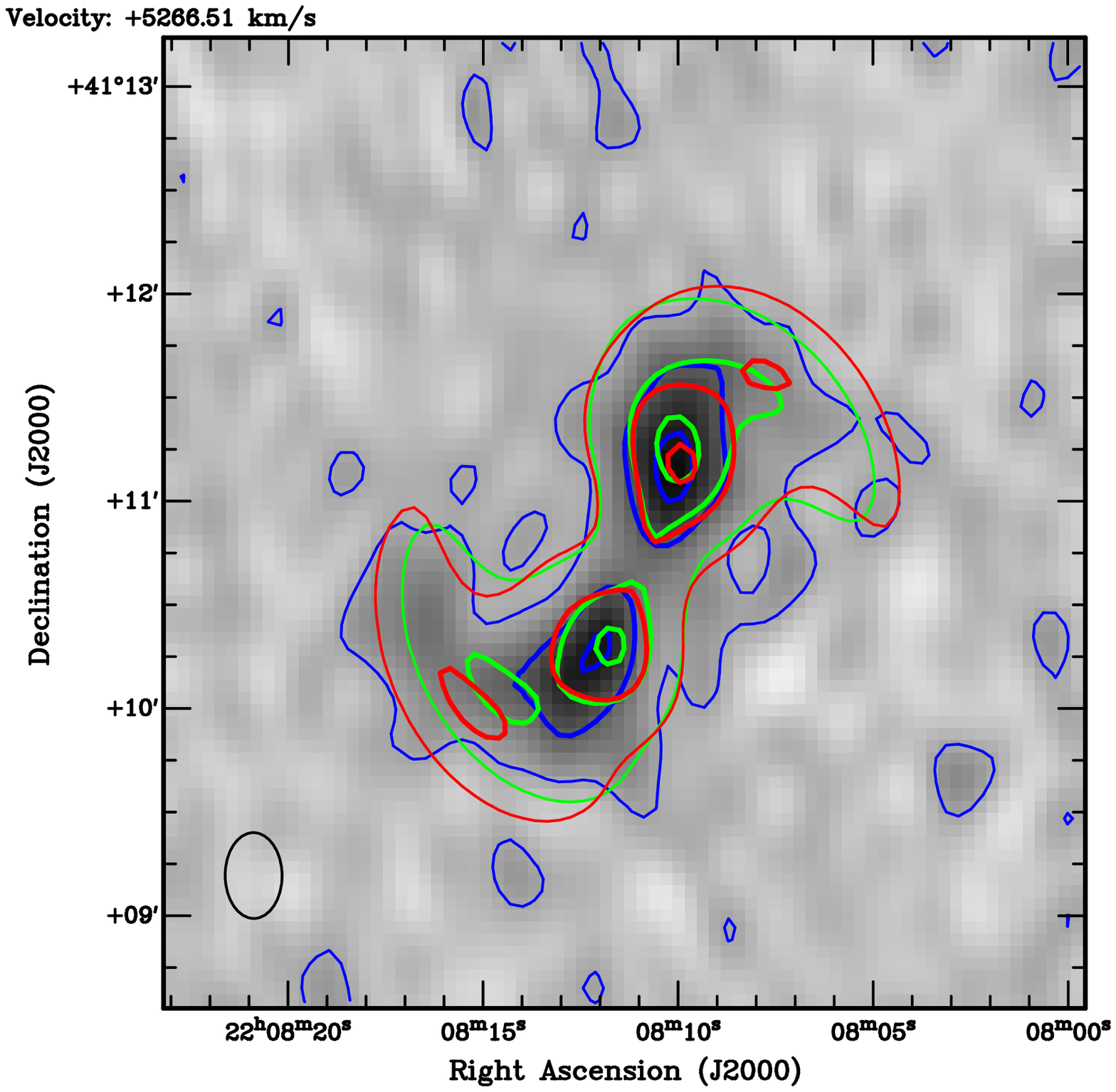}
\includegraphics[width=0.3\textwidth,keepaspectratio]{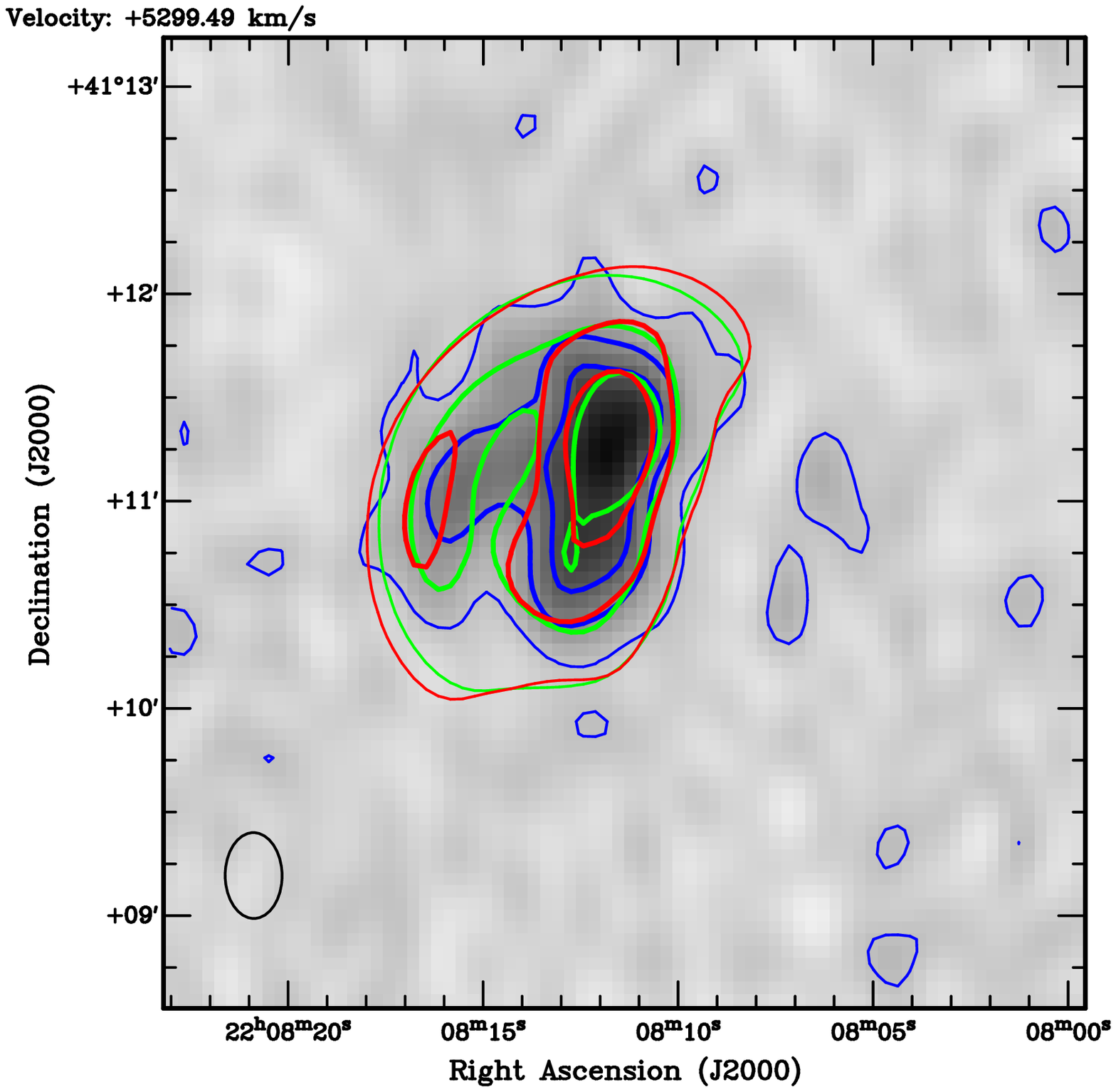}
\includegraphics[width=0.3\textwidth,keepaspectratio]{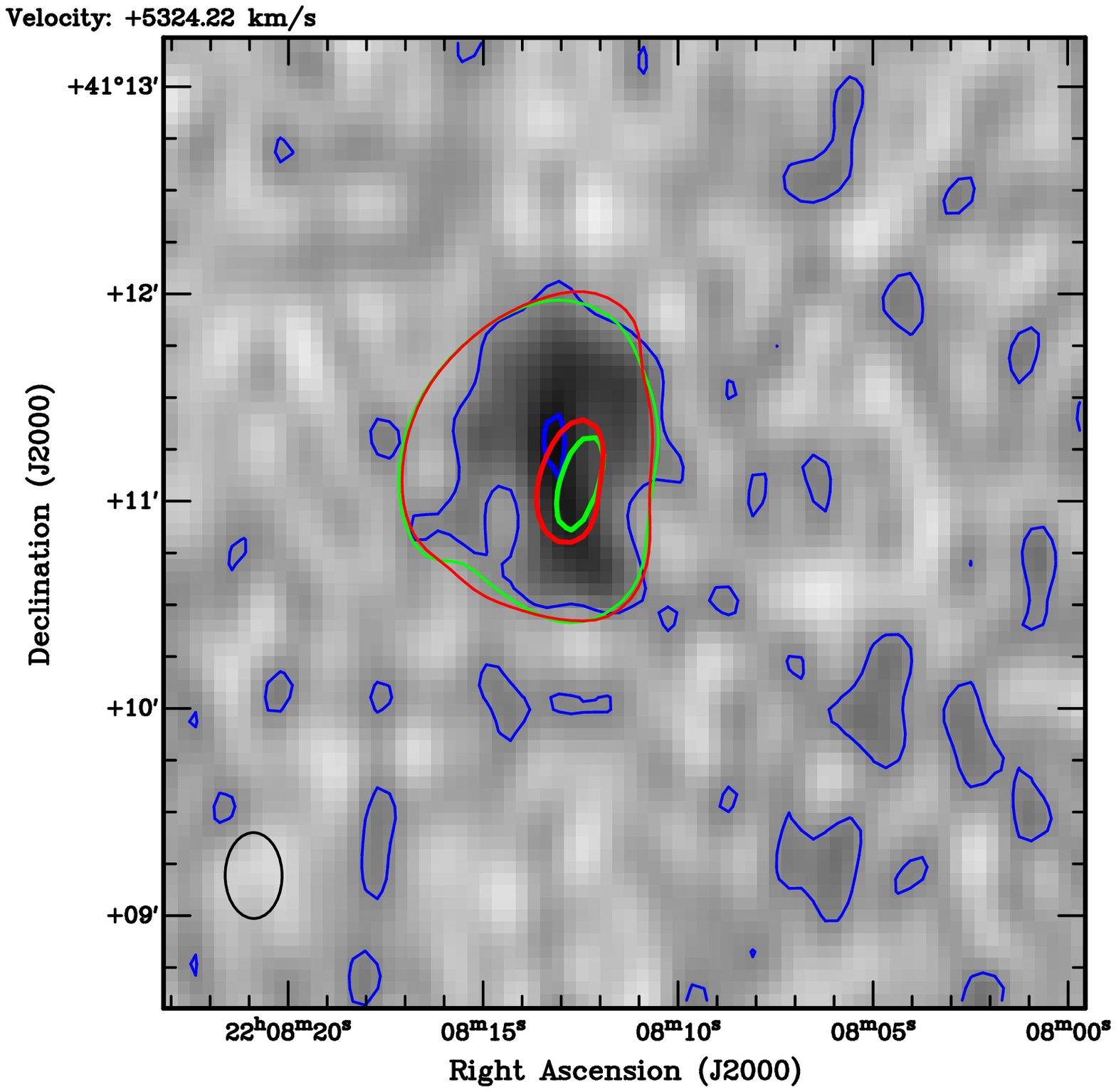}
\caption{ Selected images of H{\sc i} data cubes of UGC~11919. Blue, red, and 
green contours represent the 0.75, 3, and 5 mJy/beam levels of the observed  
and model data cubes for $i=30\degr$ and $i=13\degr$, respectively.} 
\label{datacube}
\end{figure*}

\begin{figure*}
\centering
\includegraphics[width=0.3\textwidth,keepaspectratio]{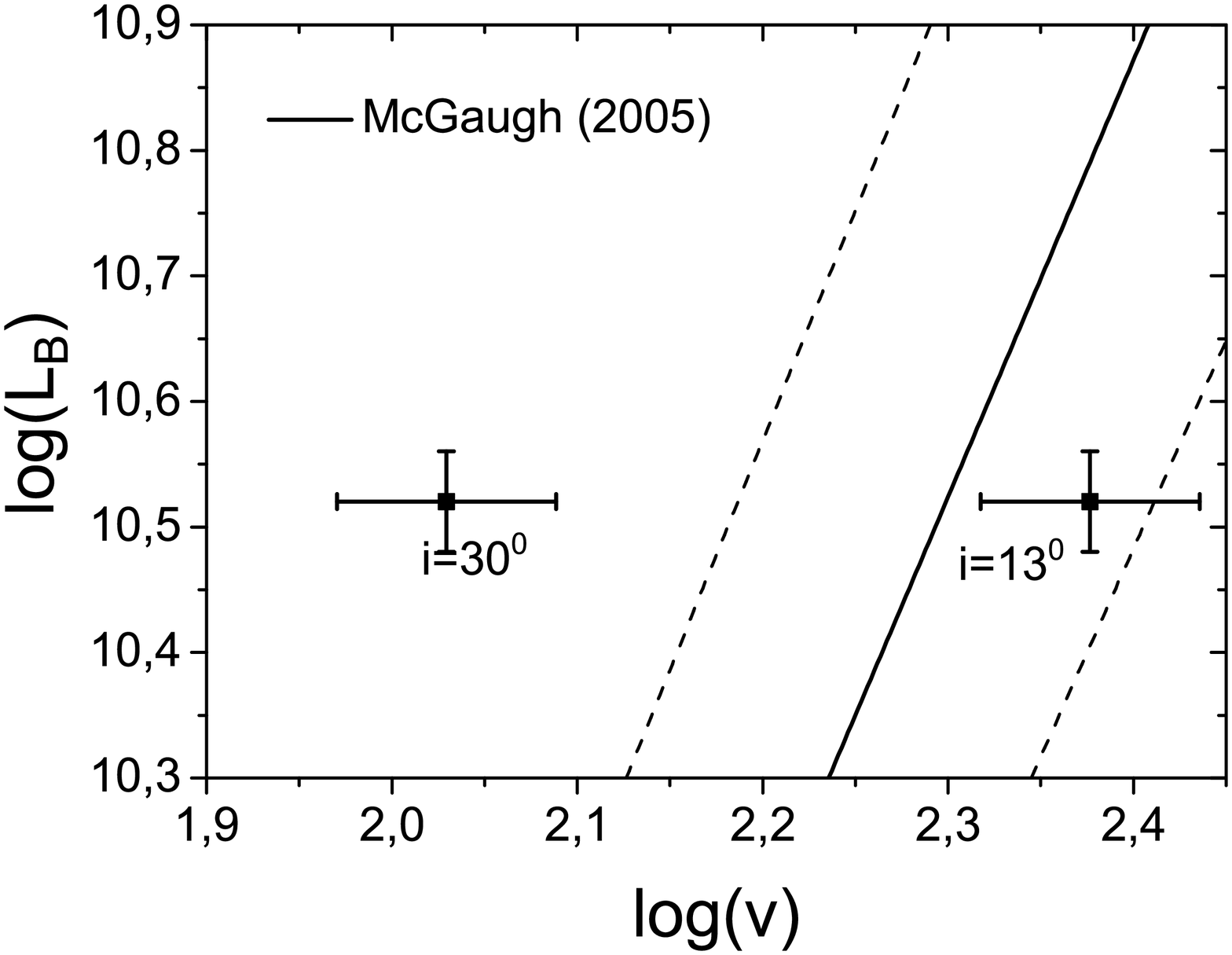}
\includegraphics[width=0.3\textwidth,keepaspectratio]{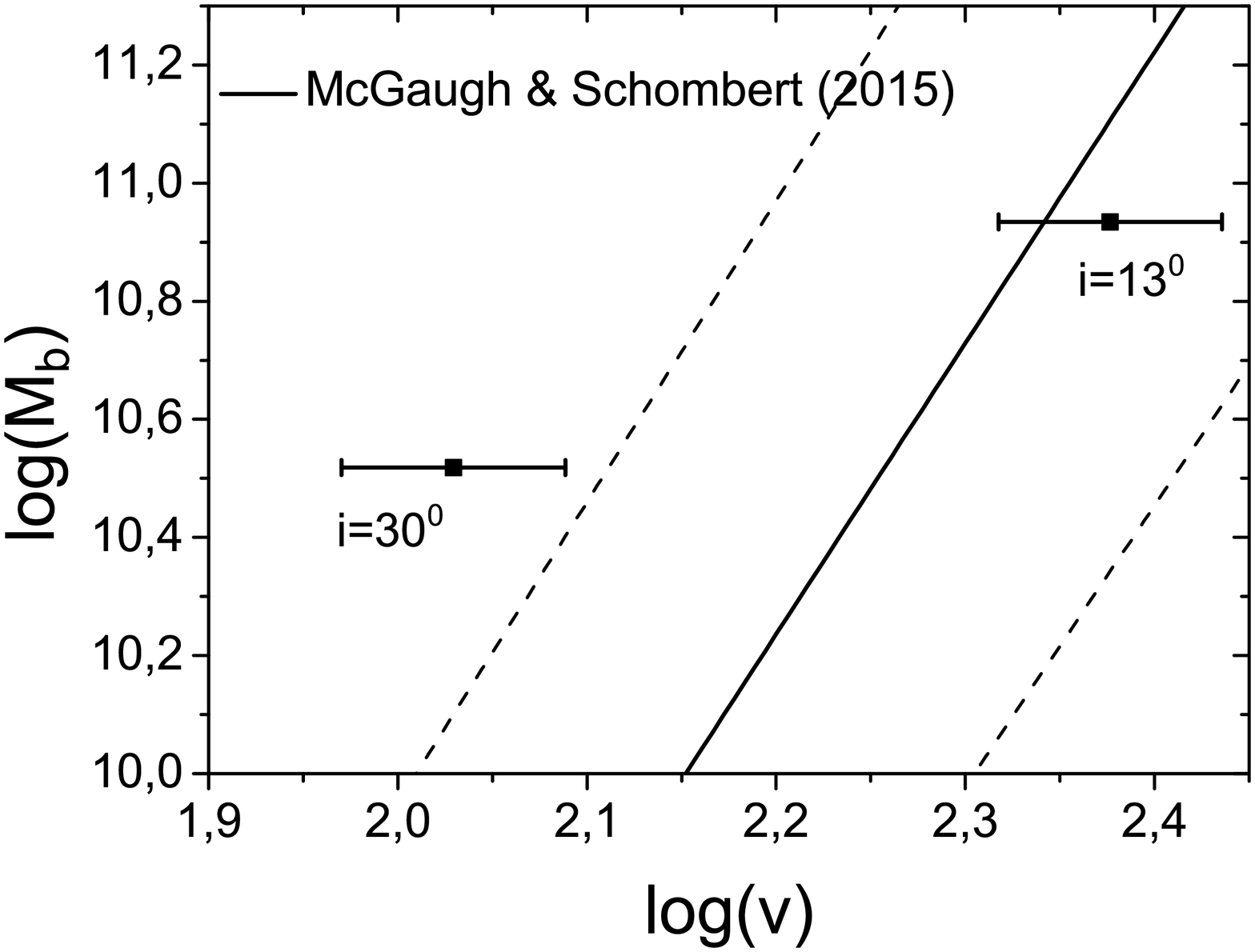}
\includegraphics[width=0.3\textwidth,keepaspectratio]{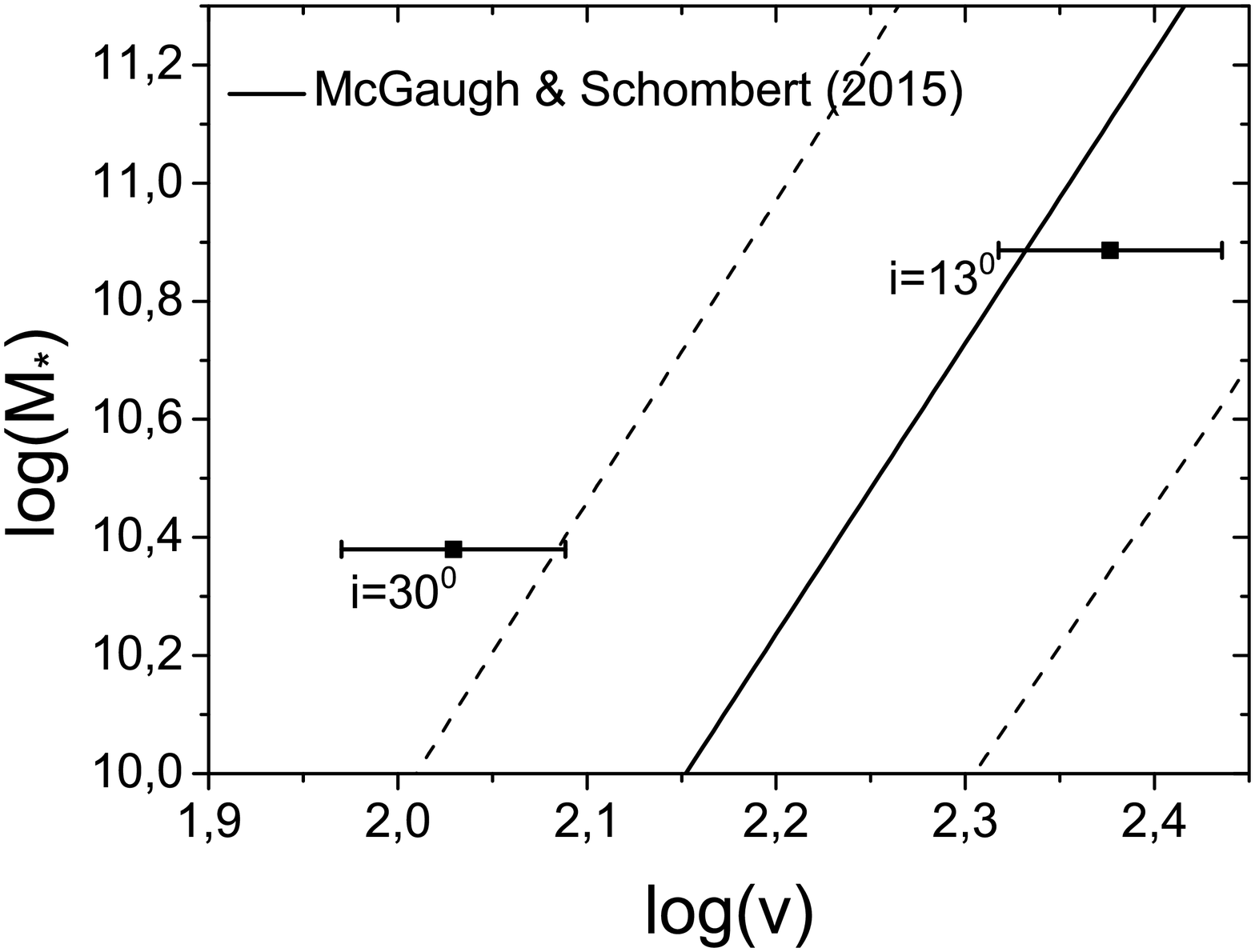}
\caption{ The position of UGC11919 for two values of inclination angle on the Tully-Fisher diagrams. Left: the 'classical' ($\log(L_B)$ versus $\log(v)$) Tully-Fisher relation, the lines correspond to the relation found by \citealt{McGaugh2005} and its uncertainty. Central and right hand panels show baryonic ($\log(M_b)$ versus $\log(v)$) and stellar ($\log(M_*)$ versus $\log(v)$)  Tully-Fisher relations respectively, the lines denote the sequences found by \citealt{McGaugh2015} and their uncertainties.  } 
\label{tf}
\end{figure*}


\section{Spectral observations and data reduction} 

We carried out observations of UGC11919 in the primary focus of the 6-m
telescope of the Special Astrophysical Observatory (SAO RAS) in August 2013
using the SCORPIO focal reducer \citep{AfanasievMoiseev2005} in the long-slit
mode. Observations were conducted for two position angles
$PA_{1}=20\degr$  which is by $25\degr$ less than the
major axis $PA_0=45\degr$ found in Paper I from the H{\sc i} data cube modeling, however it  is close to the  $PA_0$  derived from the analysis of optical isophotes, and $PA_2=-52\degr$ which is close to minor axis.  We used the two spectral cuts to study the dynamics of the disc and the spectral properties of stellar population. Unfortunately we failed to get a reliable estimates of the orientation angles of the disc from these data. We give preference to the value of $PA_0$ found from the H{\sc i} data.

The exposure times and atmospheric conditions are given
in Table \ref{tab1}. We utilized the slit-width of 1$\arcsec$.  The positions of the slit are shown in Fig. \ref{map}. We
used VPHG2300G grism with the spectral range of 4800-5570 \AA  ~and reciprocal dispersion
0.38 ~\AA$/px$. The spectral resolution is $R\approx2200$. The scale along the slit is $0.36\arcsec/px$.

The data
reduction was performed as it is described below. First we subtracted the averaged bias
frame and truncated the overscan regions. The traces of cosmic ray particles
were removed and after that all frames were divided by normalized flat field
frames. Then to calibrate the spectra we constructed two-dimensional dispersion
equation using the spectrum of a He-Ne-Ar calibration lamp, which gave the
mean error 0.05 \AA.  The spectra were linearized and summed. Then we subtracted
the spectrum of the night sky taking into account the instrumental profile
variations along the slit (for more details of sky subtraction see
\citealt{Katkov2011}). After that we performed the flux calibration using the spectrum of spectrophotometric stellar standards GD248 and GRW70D.

To estimate the variation of the parameters of the spectrograph instrumental profile we analyzed the twilight sky spectrum observed in the same observational run. To do
this, we splitted the frame of the twilight sky spectrum into the sections and co-added the spectra
to achieve a high signal-to-noise ratio. After that 
we fitted the every section by the high-
resolution solar spectrum, taken from the ELODIE3.1
stellar spectral library (\citealt{Prugniel2007}). During the fitting we parameterized the twilight sky spectrum by the Gauss-Hermite series of orthogonal functions (\citealt{vanderMarel}). It allowed us to obtain the instrumental profile parameters for different spectral ranges and along the slit. The resulting variation of the instrumental profile was taken into consideration during the spectral fitting by convolving it with the grid of stellar population models.

\begin{figure*}
\centering
\includegraphics[width=0.4\textwidth,keepaspectratio]{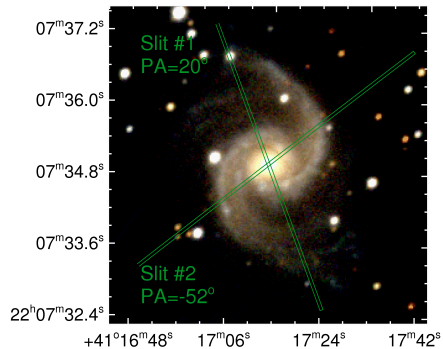}
\caption{Composite gri- image of UGC11919 from SDSS with the overplotted positions of the slits.} 
\label{map}
\end{figure*}

The parameters of stellar population of galaxy were obtained by fitting 
high-resolution PEGASE.HR simple stellar population models \citep{LeBorgneetal2004}  (hereafter SSP models) in which all stars are formed in one instantaneous burst of starformation. 
 The NBursts full spectral fitting technique
\citep{Chilingarian2007} we used allows to fit the observed
spectrum in the pixel space against the population model convolved with a parametric line-of-sight velocity distribution. In this method the parameters of the stellar populations are derived by nonlinear minimization of the quadratic difference chi-square between the observed and model spectra. As the result of modelling we
  obtained the radial profiles of the velocity, velocity dispersion, age of stellar population and its metallicity. A subtraction of the model of stellar spectrum from the observed one gave us a pure emission spectrum. We fitted the emission lines by Gaussian distribution which allowed to obtain the velocity and velocity dispersion of the ionized gas.

In order to increase the signal to noise ratio for the peripheral regions of the galaxy we used  the adaptive binning procedure to achieve the signal-to-noise ratio $S/N=10$ for every bin.

As far as the mass-to-light ratio of different components of UGC11919 is of crucial interest for us we have also performed a more elaborate analysis to obtain the integrated properties of stellar population of the main stellar components of the galaxy. In contrast to the estimation of radial distribution of the parameters of stellar population described above we utilized the NBursts+phot technique presented by \citet{Chilingarian2012}.  This technique uses not only the spectral data but also the broad- band spectral energy
distribution  (SED). This method was earlier successfully applied to the giant low surface brightness galaxy Malin~2 (see \citealt{Kasparova2014}). We used the
photometric broad-band optical data from Paper I parallel with the archive
SDSS and GALEX images in NUV, FUV, u, g, r, i, z, B, V photometric bands. The
photometric data were reduced in a standard way using the MIDAS software package\footnote{
MIDAS is developed and maintained by the European Southern Observatory.
software package. The archive data were calibrated according to the
information available at the FITS file headers.}.  After that we modelled the
spectra and SEDs separately for the disc, bulge and nuclear disc regions assigning the equal  weight to the
 wide-band magnitudes and the detailed spectrum fitting. In contrast to the estimation of radial distribution of the parameters of stellar population described above here we used the integrated spectrum for each component of the galaxy. To do it we summed the spectra in the following three regions: $r<3$\arcsec (nuclear disc), $4\arcsec<r<8$\arcsec (bulge), $r>9$\arcsec (the main disc). To fit both  the
spectra and SED we used not only the SSP model as for the radial variation of the parameters, but also the models with 
the exponentially declined star formation 
(exp-SFH). Both SSP and exp-SFH models are computed with the
PEGASE.HR code based on the~ELODIE.3.1 empirical stellar library. The SSP
stellar population models are characterized by metallicity [Z/H] and age $T$, the~exp-
SFH models~--- by metallicity [Z/H] and the exponential decay time scale $\tau$.
The epoch of start of star formation in the exp-SFH model is fixed at 
$T=13$ Gyr (note that the results are not sensitive to the choice of $T >10^{10}$ yr).  We used two IMFs in the models: the Salpeter IMF  (SSP
model) and the Kroupa IMF  (exp-SFH model). 
\begin{table}
\begin{center}
\caption{Observation log. \label{tab1}}
\begin{tabular}{lll}
\hline
$PA, \degr$ & Exposure time, s & Seeing, $\arcsec$\\
20  & 8100 & 1.4\\
-52 & 9600 & 1.4\\
\hline
\end{tabular}
\end{center}
\end{table}

\section{Results}

\subsection{Kinematics}

In Figs. \ref{kinprofiles20},\ref{kinprofiles52} we show kinematical profiles
of line-of-sight velocities and velocity dispersions of stars and ionized gas obtained from the long-
slit spectra oriented at $PA_1=20\degr$
  and $PA_2=-52\degr$. We also give the reference composite SDSS images in gri- bands in the top panels.

Line-of-sight velocity distribution
along $PA_1$ clearly reveals the rotation with a steep velocity
gradient within galactocentric distance $r\approx 5$\arcsec. 


\begin{figure}
\centering
\includegraphics[width=0.4\textwidth,keepaspectratio]{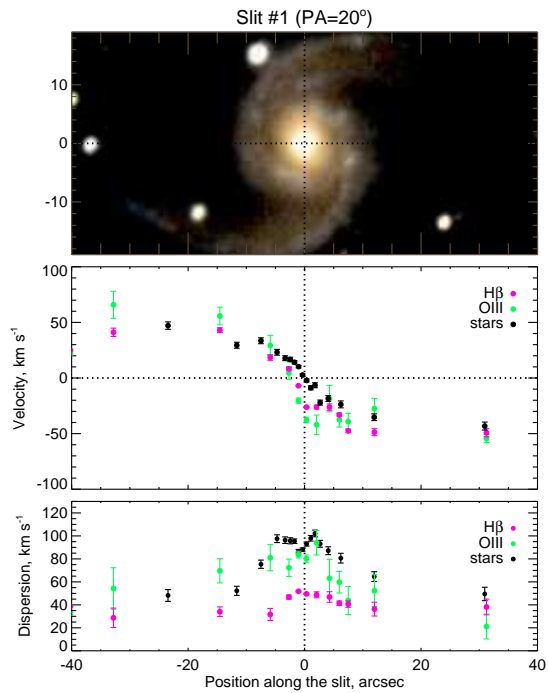}
\caption{The kinematic profiles of the line-of-sight velocities of stars and 
ionized gas along $PA_1=20\degr$. Top panel corresponds to the SDSS composite gri- reference image of UGC11919. Central and bottom panels show the line-of-sight velocity and velocity dispersion profiles correspondingly.} 
\label{kinprofiles20}
\end{figure}

Velocity dispersion of stars is high in the central 5-7\arcsec, where the light of bulge prevails.   From Figs. \ref{kinprofiles20},\ref{kinprofiles52} it follows
that there is a central depression of stellar velocity dispersion for both
slit positions, within $r \approx 4\arcsec$ (corresponding to $r \sim 1.4~kpc$)
which can indicate a nuclear, kinematically decoupled, stellar disc.  This effect could also be the result of the orbital anisotropy  related to the bar, however in this case we would expect it to be different for nearly orthogonal positions of the slit. 

\begin{figure}
\centering
\includegraphics[width=0.4\textwidth,keepaspectratio]{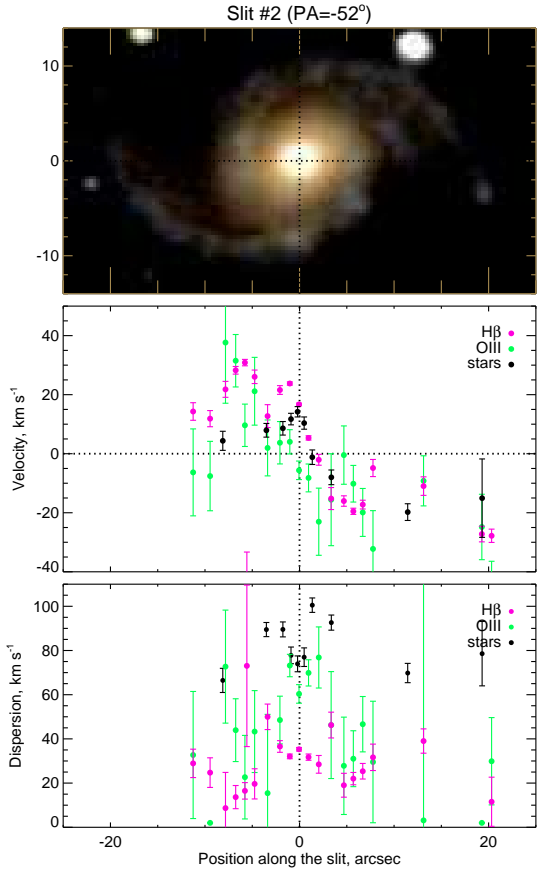}
\caption{The kinematical profiles of the line-of-sight velocities of stars 
and ionized gas along $PA_2=-52\degr$. Top panel corresponds to the SDSS composite gri- reference image of UGC11919. Central and bottom panels show the line-of-sight velocity and velocity dispersion profiles correspondingly.  } 
\label{kinprofiles52}
\end{figure}

\begin{figure}
\centering
\includegraphics[width=0.4\textwidth,keepaspectratio]{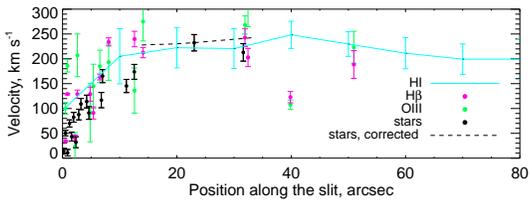}
\caption{The rotation curve of UGC11919 assuming $i=13 \degr$. Black circles correspond to the radial profile of the velocity of stars. Green  and pink  circles show the rotation curve of ionized gas. Cyan line corresponds to the H{\sc i}  rotation curve. Dashed black line demonstrates the rotation velocity of stars corrected for asymmetric drift. } 
\label{rc}
\end{figure}

In Fig. \ref{rc} we show the rotation curve of stars and gas assuming $i=13 \degr$ (the long-slit data are taken along $PA_1=20\degr$). We take into account that the slit orientation differs by the angle $\alpha$ from the position of major axis, using the geometric relationships for axisymmetric disc:
\begin{equation}
\label{formula_v} v(r) = \frac{v_r(r)\sqrt{(sec^2(i)-\tan^2(i)\cos^2(\alpha))}}{\sin(i) \cos(\alpha)}
\end{equation}

\begin{equation}
\label{formula_r} r = r_\alpha \sqrt{(\sec^2(i) - \tan^2(i) \cos^2 (\alpha))}
\end{equation}
Here  $ r_\alpha $ is the radius in the plane of projection, $v_r$ is the line-of-sight velocity corrected for the systemic velocity. 

In Fig. \ref{rc} we also show by black dashed line the rotation velocity of stars corrected for asymmetric drift in epicyclical approximation for exponential stellar disc following \cite{BinneyTremaine}:
\begin{equation}
\label{formula_drift}v_c^2 = v_r^2 + c_r^2\cdot  (0.5 \frac{d\ln(v_r)}{d\ln(r)} -0.5 +\frac{r}{r_d} -  \frac{d \ln (c_r^2)}{d\ln(r)})
\end{equation}
Here $v_c$ is the circular velocity, $v_r$ is the observed rotation velocity of stars, $c_r$ is radial velocity dispersion of stars, $r_d$ is the exponential scalelength of disc.

As could be seen from Fig.
\ref{rc}  the optical rotation
velocities agree well with the H{\sc i} kinematic data, with the exception of the central part of the disc where the resolution of H{\sc i} is not good enough.   The low intensity of emission lines  cannot
reproduce the reliable shape of rotation curve in the outer regions of the disc. By this reason we'll use below
the rotation curve found from H{\sc i} measurements, which allow to
trace it much further from the centre.

\subsection{The properties of stellar population}
Parallel with kinematic estimates, we also obtained the radial profiles of luminosity-weighted values of stellar age $T$ and metallicity [Z/H]. The radial profiles of the resulting parameters are shown in Figs. \ref{popprofiles20}, \ref{popprofiles52} for $PA_1$ and $PA_2$ respectively. We also give the reference SDSS images in order to link the data with the image of the galaxy. As could be seen from Figs. \ref{popprofiles20}, \ref{popprofiles52} the age and metallicity do not remain constant along the slits. We mark by thin dotted lines the positions of regions with sharp gradients of the age of stellar population. The central depression of the age noticeable in the profiles for both slit positions most probably corresponds to the nuclear disc with the long lasted star formation which reduces the mean stellar age. The increase of the age at $r\approx 3-4$\arcsec seen in Figs. \ref{popprofiles20}, \ref{popprofiles52} is the manifestation of the old stellar bulge, which dominates the luminosity in this region.  We also see in Fig. \ref{popprofiles20} two minima at the positions close to the spiral arms, evidently indicating the presence of the current star formation in the spiral arms.

\begin{figure}
\centering
\includegraphics[width=0.4\textwidth,keepaspectratio]{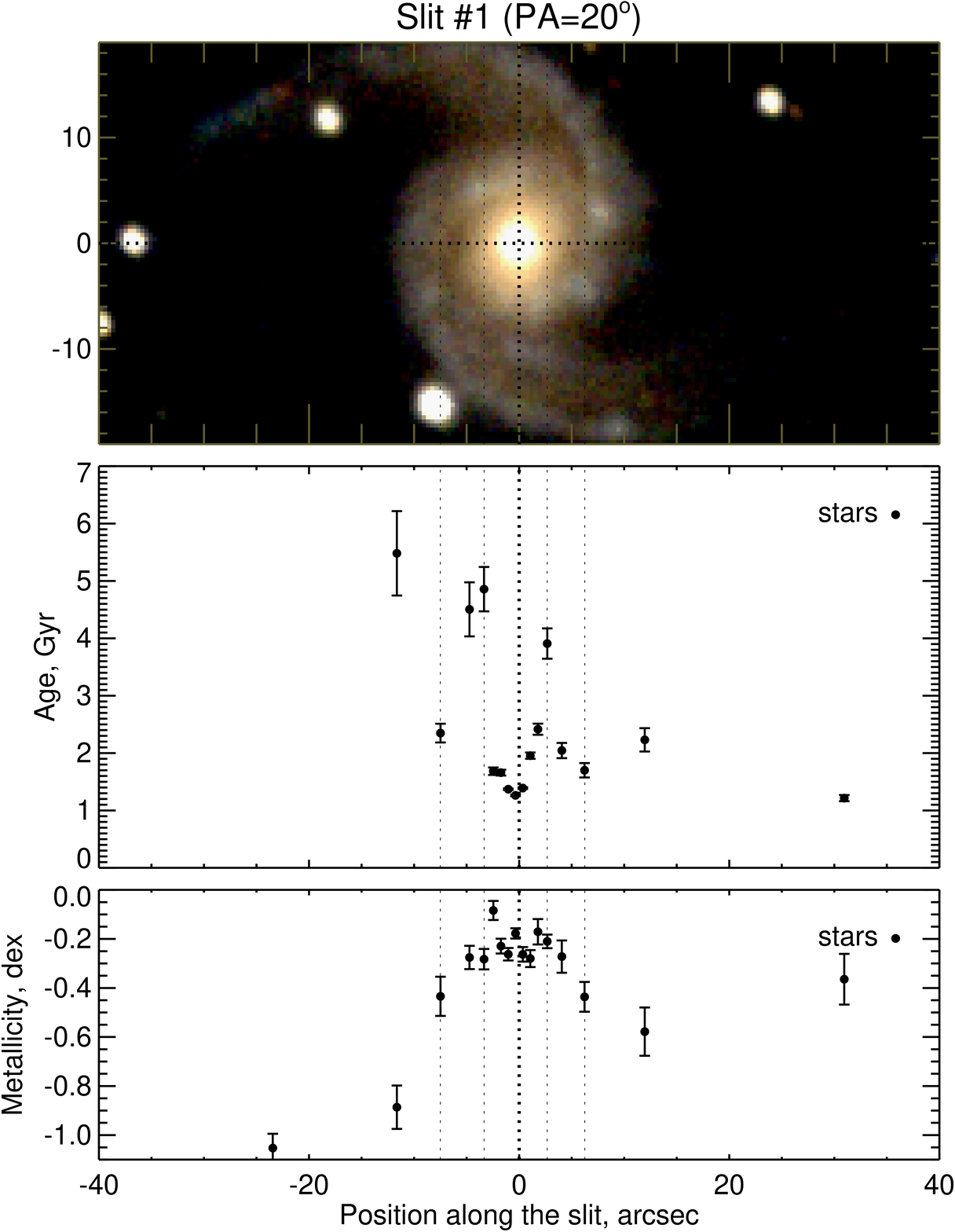}
\caption{Stellar population age and metallicity radial profiles along 
$PA_1=20\degr$ (middle and bottom panels). Top panel corresponds to the SDSS composite gri- reference image of UGC11919. Dotted thin vertical lines mark the position of the regions where the age of stellar population changes sharply. } 
\label{popprofiles20}
\end{figure}

\begin{figure}
\centering
\includegraphics[width=0.4\textwidth,keepaspectratio]{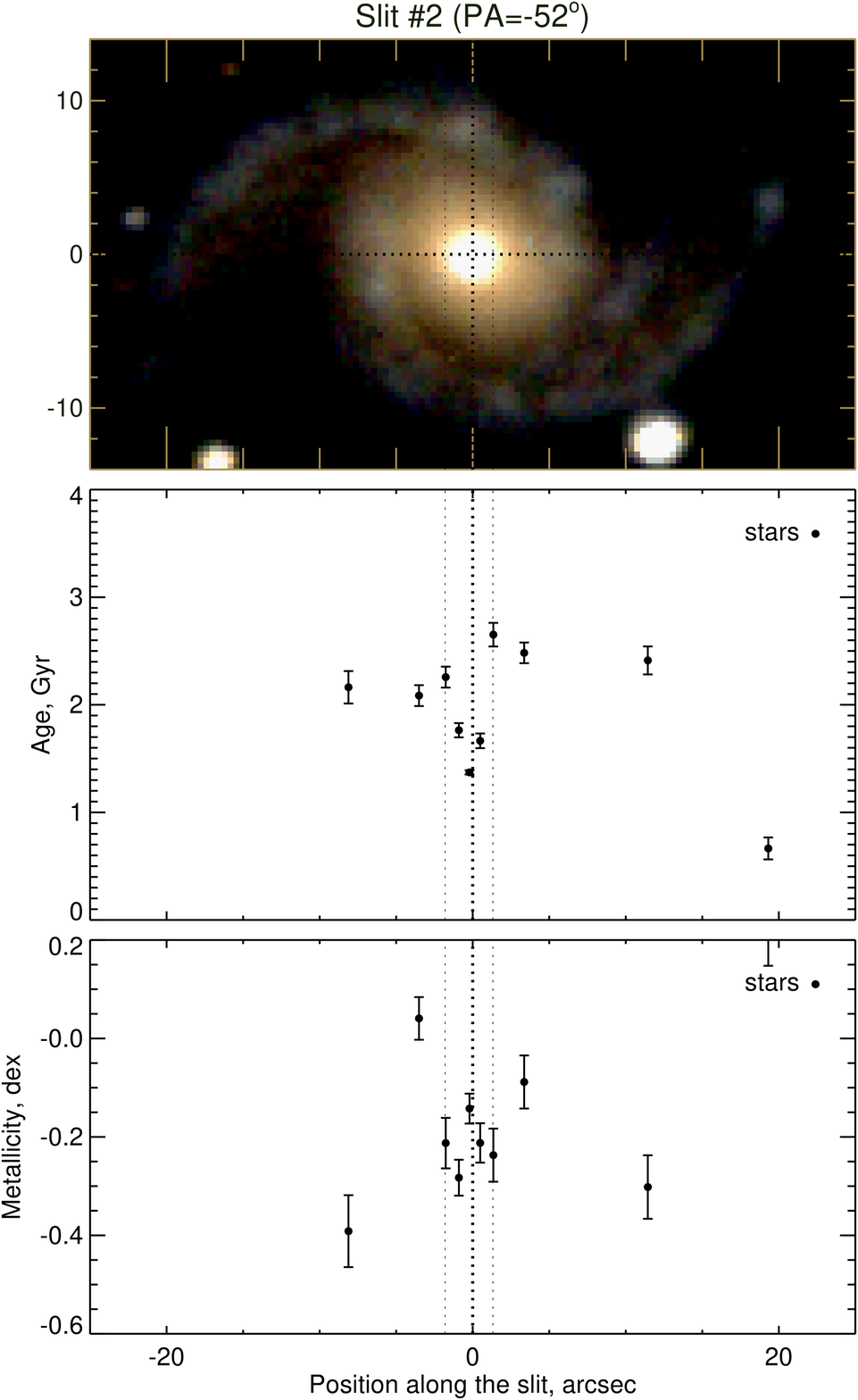}
\caption{Stellar population age and metallicity along $PA_2=-52\degr$ (middle and bottom panels). Top panel corresponds to the SDSS composite gri- reference image of UGC11919. Dotted thin vertical lines mark the position of the regions where the age of stellar population changes sharply.} 
\label{popprofiles52}
\end{figure}

The mean (luminosity-weighted) stellar age obtained for the central region $r< 4\arcsec$ of UGC11919 is about 1.5 Gyr. Stellar metallicity of this region is about -0.2 dex, that is a factor 1.5 lower than the solar metallicity. This underabundance is typical for a galaxy of a given luminosity.  
Further from the centre where the disc contribution dominates we obtained the mean values: $[Z/H]\approx -0.4$, $T\approx 2.2$ Gyr
corresponding to $M/L_B=1.6$ which 
practically coincides with
that found by Paper I from the observed color index $(B-V)_0$ and
$M/L_B$-color relation given by \cite{bdj} ($(M/L_B)_{disc}=1.7\pm 0.3$). This ratio is about 3 times higher
than that based on the mass-modeling for $i = 30 \degr$ and is in good agreement with the rotation curve for $i=13 \degr$ (see Sect. \ref{mm}).

\subsection{SED and the detailed spectrum modelling}\label{ss}

To obtain the integrated properties of stellar population of each component of the galaxy we performed modeling of both spectrum and broad-band SED (for more details on the method see Sect. 3).

As could be seen from Fig. \ref{discssp2}, SSP models failed to reproduce the
high fluxes in UV. However the GALEX images of UGC11919 are characterized by low S/N
ratio. By that reason we decided to consider the SSP models neglecting  UV
fluxes. In Figs. \ref{discssp}, \ref{bulgessp} and \ref {idssp} we show the
observed and model spectra (SSP model and SED model) for the disc, bulge and nuclear disc
correspondingly without using the UV data. Blue and black lines and the symbols correspond to the
model and observed data respectively. The resulting values of stellar population age, metallicity, and $M/L_B$ ratios for the SSP modeling of the spectra obtained for both slits are given in Table \ref{tab2}. 
The obtained mass-to-light ratio (see Table \ref{tab2}) is 1.4 times higher and 1.1 times lower than that found in Paper I from $(B-V)_0$ color indices for a bulge and a disc correspondingly. All these values of
$M/L_B$ ratios remain significantly higher than those obtained in mass-
modeling for $i=30$\degr.  
However for $i=13 \degr$ the discrepancy disappears. It  indicates
that either the IMF of UGC11919 is  bottom-light
with a deficit of low massive stars, or the inclination angle is much lower than the photometry gives.

\begin{figure*}
\centering
\includegraphics[width=\textwidth,keepaspectratio]{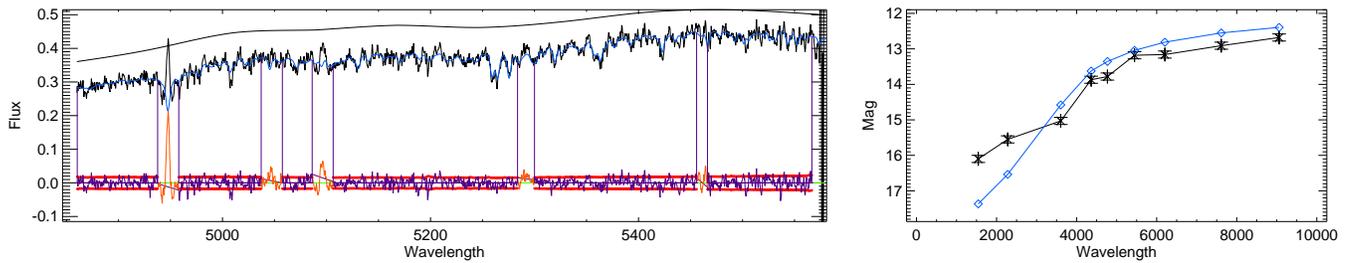}
\caption{The results of modeling of the integrated disc spectrum at $PA_1=20\degr$ 
(left) and disc SED (right) for SSP model. Blue and black lines and symbols 
correspond to the model and observed data respectively.
} 
\label{discssp2}
\end{figure*}

\begin{figure*}
\centering
\includegraphics[width=\textwidth,keepaspectratio]{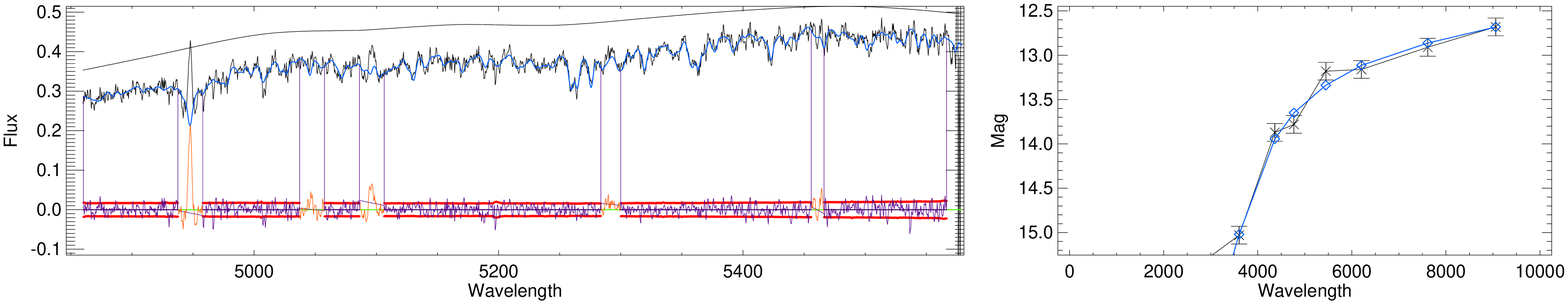}
\caption{The results of modeling of the integrated disc spectrum at $PA_1=20\degr$ 
(left) and disc SED (right) for SSP model without taking into account the UV data. Blue and black lines and symbols 
correspond to the model and observed data respectively.
} 
\label{discssp}
\end{figure*}

\begin{figure*}
\centering
\includegraphics[width=\textwidth,keepaspectratio]{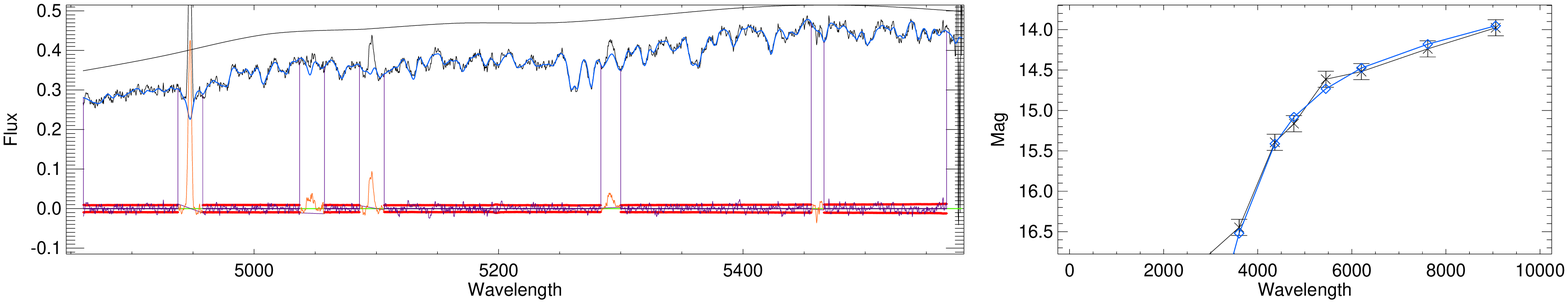}
\caption{The results of modeling of the integrated  bulge ($3<r<8$\arcsec) 
spectrum at $PA_1=20\degr$ (left) and the restored SED (right) for SSP model without taking into account the UV data. 
Blue and black lines and symbols correspond to the model and observed 
data respectively.
} 
\label{bulgessp}
\end{figure*}

\begin{figure*}
\centering
\includegraphics[width=\textwidth,keepaspectratio]{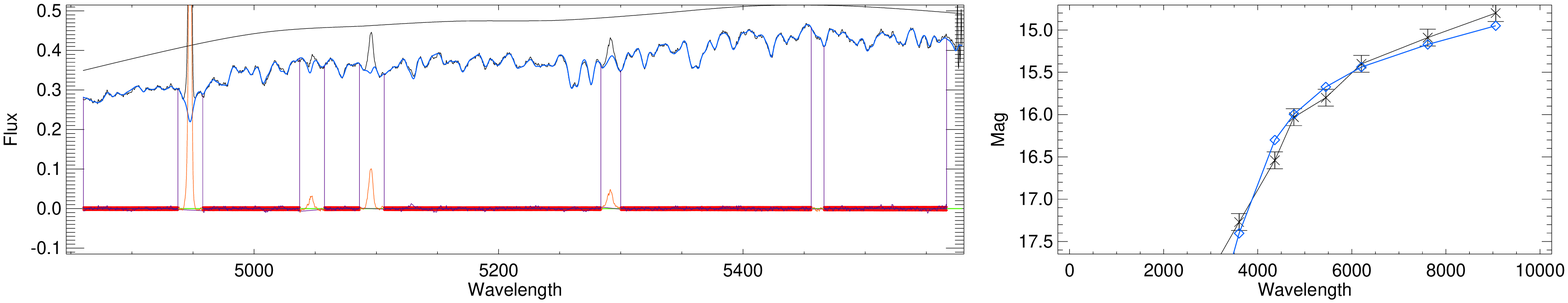}
\caption{The results of modeling of the integrated  inner disc ($r<3$\arcsec) 
spectrum at $PA_1=20\degr$ (left) and the restored SED (right) for SSP model without taking into account the UV data. 
Blue  and black lines and symbols correspond to the model and observed 
data respectively.
} 
\label{idssp}
\end{figure*}

Figs. \ref{discexp} and \ref{bulgeexp} show the observed and exp-SFH model
spectra and SED for disc and bulge. The model with exponential star formation history failed to reproduce the spectrum and SED of the nuclear disc. It means that a history of its star formation should be more complex. Lines and symbols
correspond to the model (blue) and observed (black) data.  The resulting parameters $\tau$, $Z$ and $M/L_B$ for the spectra obtained at both slit positions using exp-SFH model are given in Table \ref{tab2}.

As could be seen from Table \ref{tab2}, there is a reasonable agreement between the parameters obtained for the spectra of both slits. The variation of the age and metallicity for different $PA$s are quite expectable because the slits cross the regions with different properties of stellar population. From Table \ref{tab2} it follows that the youngest stellar component of the galaxy is the nuclear disc, at the same time it possesses the relatively low metallicity. One may propose that this component could be formed by the accretion of the low mass satellite galaxy by UGC11919 followed by the star formation. The stellar population of bulge possesses the largest age,  its metallicity is higher than that of the younger stellar main disc reflecting the negative radial metallicity gradient usually observed for spiral galaxies.

\subsection{On the disc thickness and  its gravitational stability }

The measurement of stellar velocity dispersion within one disc radial scalelength $r_d\approx 21\arcsec \approx 7.5$ kpc where the contribution of bulge is negligible  may give the additional information about the disc dynamical condition  for the different  accepted disc inclination. 

In both cases - for  $i=30 \degr$ or $i=13 \degr$  the main component of stellar velocity dispersion is the vertical component $c_z$ perpendicular to the disc plane. It allows to use its value (~48 km $s^{-1}$) for the rough estimate of the equilibrium scaleheight $Z_0$ of stellar disc:
\begin{equation}
Z_0 = c_z^2/\pi G\sigma_d(r),
\end{equation} 
where $\sigma_d(r)$ is the surface density of the disc at the galactocentric distance r. Mass decomposition of the galaxy presented in Paper I gives for $r=r_d$ the density $\sigma_d(r_d) \approx 33 M_\odot/pc^2$ for $i=30 \degr$ (disc with peculiarly low $M/L$ ratio). For the inclination $i=13 \degr$, expected for the normal stellar IMF   $\sigma_d(r_d) \approx 112 M_\odot/pc^2$. The corresponding values of $Z_0$ are 4.9 kpc and 1.5 kpc or $Z_0/r_d=0.7$ and $Z_0/r_d=0.2$. It is evident that in the first case the disc half-thickness is unacceptably big, which gives the additional evidence in favor of the lower inclination and the normal density stellar disc.}

The estimates of stellar velocity dispersion may also be used to put constraint on the disc surface density using the condition of gravitational stability of the disc. It is worth noting that the comparison of the masses of galactic discs of spiral galaxies found under assumption of their marginal stability and those found from photometric models shows that they usually agree, which gives
evidences that the stellar discs are close to marginally
stable condition at least at 1-2 radial scalelengths
 (see e.g. \citealt{Zasov2004},
\citealt{Zasov2011}, \citealt{Saburova2012}).
For the one-component isothermal disc the stable state is locally reached
when the radial stellar velocity dispersion $c_r$ at a given galactocentric distance r
exceeds the critical value :

\begin{equation}
c_{r~crit.}=Q_T\cdot 3.36G \sigma_d / \varkappa,  
\end{equation} 
where $\varkappa$ is the epicyclic frequency, and $Q_T$ is the Toomre' stability  parameter which is equal to unit for pure radial perturbations of a thin disc. Numerical simulations show that for the marginal stability of exponential discs with finite thickness the parameter  $Q_T\approx$ 1.2 - 3 is
 slowly growing to the disc periphery (see f.e. \citealt{Khoperskov2003}). In general case, one have to take into account the presence of the additional dynamically cold component of the galaxy (gas) which makes the disc more unstable (see for example \citealt{Romeo2011}). However in our case the local values of $\sigma_{gas}$ do not exceed 9 $M_\odot/pc^2$ (see Paper I), whereas the density $\sigma_d $ obtained from the rotation curve for $r = r_d$ is at least one order of magnitudes higher. It means that the presence of a gas component with a typical velocity dispersion ~ 10 $km~s^{-1}$ does not change the stability condition significantly especially if to take into account the approximate nature of our estimates. 
  We choose again a representing galactocentric  distance $r_d\approx 21\arcsec$
(one radial scalelength), where in addition to the velocity of rotation we also have the estimates of stellar velocity dispersion.

The marginal velocity dispersion $c_r$ depends on the
assumed disc inclination. Indeed, the reducing of $i$ will increase both the
value of the surface density $\sigma$ found from the rotation curve
decomposition,  and $\varkappa$, that is the numerator and denominator in the equation above.
However $\varkappa$ at a given $r$ is proportional to the velocity of rotation $v$ which, in turn, is proportional to $1/\sin i$, whereas $\sigma(r_d) \sim v^2\sim 1/\sin^2i$. It enables to
compare the models with different inclination angles $i$.

We estimated
radial velocity dispersion from the observed line-of-sight stellar velocity
dispersion $c_{obs}$ at $PA_1$,  taking into account the expected links
between the dispersion along the radial, azimuthal and vertical directions:

$$ 
\begin{array} {l}
$$c_{obs}^2(r) = (c_z^2 \cdot \cos^2 (i)+ c_{\phi}^2\cdot \sin^2 (i)\cdot \cos^2(\alpha) +\\ 
c_{r}\sin^2(i)\cdot \sin^2(\alpha)$$
 \end{array}
$$

where $\alpha$  is the angle between the direction of the slit and the major axis.

To solve the equation we need two additional conditions: $ c_{r} = 2\Omega
\cdot  c_{\phi} /\varkappa $ ( Lindblad formula for the epicyclic approximation) and $ c_z =k\cdot c_{r} $, where
$c_z$, $c_{\phi}$, $c_{r}$ are the dispersion along the vertical, azimuthal
and radial directions, projected on the plane of the galaxy. The coefficient
$k$  was taken to be 0.6 in accordance with direct measurements,
which show that it could lie in the range 0.5--0.8 (see e.g.
\citealt{Shapiro2003}). The epicyclic frequency was calculated from the H{\sc
i}  rotation curve for a given inclination angle: $\varkappa(r)=2v(r)/r\sqrt{0.5+r/2v(r)( \frac{\partial v(r)}{\partial r})}$. For the accepted $i=13\degr$
we obtained $c_r=72~ \textrm{km~s}^{-1}$. The gravitational stability criterion gives 
the upper limit of the surface density of stable disc at the radial
distance of one scalelength $r_d$ as  $\approx 77
M_{\odot}/pc^2$ which is $\sim 2.3$ times higher  than it follows from the decomposition of rotation curve constructed for $i = 30\degr$ (see Paper I). However this density is comparable with the photometrically determined density at $r=r_d$, obtained in Paper I for the usually accepted stellar IMF ($\approx 112
M_{\odot}/pc^2$). If we take a more probable value $i = 13\degr$, the velocity of rotation will become about two times higher, and the marginal surface density at $r=r_d$ would be about $(2\pm0.3)\cdot 10^2 M_{\odot}/pc^2$. This value exceeds both the photometrical density estimate and the maximum density of the disc compatible with the observed rotation curve. We conclude that the disc at this distance is stable and moderately overheated. This overheating may be a result of tidal interaction or minor merging which looks quite possible  because there are several faint galaxies observed in the close vicinity from  UGC11919,  revealing themselves in the H{\sc i} map (see Paper I).
%
%

\begin{figure*}
\centering
\includegraphics[width=\textwidth,keepaspectratio]{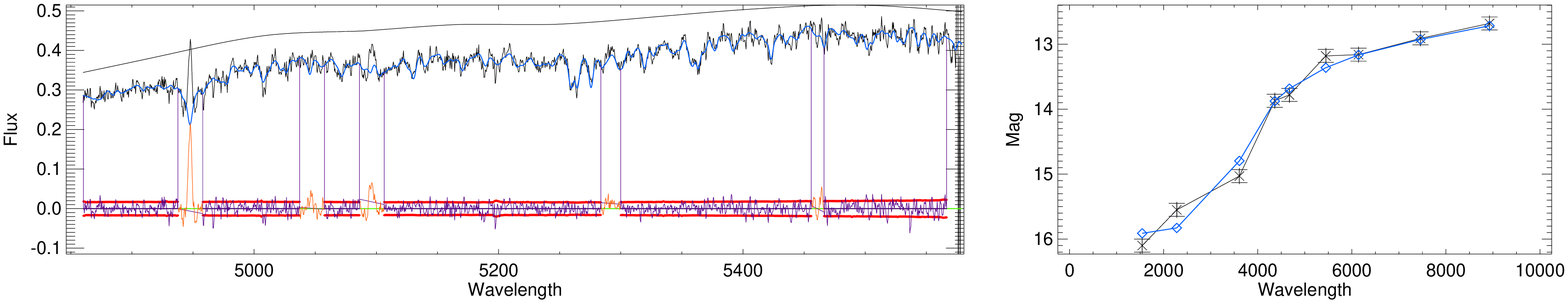}
\caption{The results of modeling of the summed disc spectrum at 
$PA_1=20\degr$ (left) and disc SED (right) for exp-SFH model. Blue  and black 
lines and symbols correspond to the model and observed data respectively.
} 
\label{discexp}
\end{figure*}

\begin{figure*}
\centering
\includegraphics[width=\textwidth,keepaspectratio]{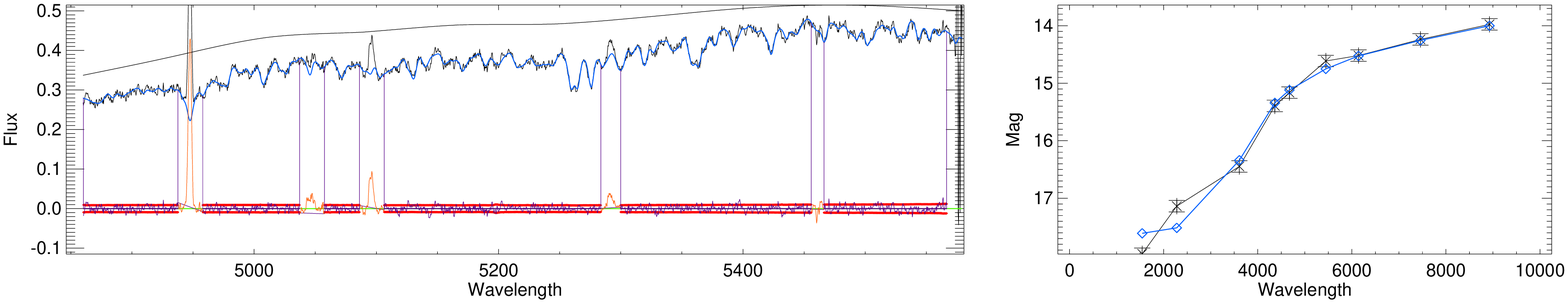}
\caption{The results of modeling of the summed bulge spectrum at $PA_1=20\degr$ 
(left) and bulge SED (right) for exp-SFH model. Blue and black lines and symbols 
correspond to the model and observed data respectively.
} 
\label{bulgeexp}
\end{figure*}

\begin{table*}
 \begin{tabular}{|c|cccccccccccc}
    \hline
&\multicolumn{6}{c|}{PA=20}&\multicolumn{6}{c}{PA=-52}\\
Model&\multicolumn{2}{c}{$T$ /\ $\tau$, Gyr}&\multicolumn{2}{c}{[Z/H], dex}&\multicolumn{2}{c}{$M/L_B$}&\multicolumn{2}{c}{$T$ /\ $\tau$, Gyr}&\multicolumn{2}{c}{[Z/H], dex}&\multicolumn{2}{c}{$M/L_B$}\\
&SSP&exp SFH& SSP&exp SFH & SSP &exp SFH & SSP&exp SFH  & SSP&exp SFH & SSP&exp SFH \\
\hline
inner disc&2.05&-&-0.29&-&1.56&-&2.48&-&-0.31&-&1.87&-\\
bulge&4&5.22&-0.44&-0.36&2.73&2.07&4.44&5.14&-0.45&-0.37&2.95&2.09\\
disc&2.55&7.75&-0.64&-0.54&1.48&1.42&2.65&7.83&-0.50&-0.37&1.69&1.56\\

\hline
 \end{tabular}
 \caption{The resulting parameters obtained by SED and spectra fitting for SSP and exp SFH models (see the text). \label{tab2}}
\end{table*}
\subsection{Mass modeling}\label{mm}
\begin{figure}
\centering
\includegraphics[width=0.4\textwidth,keepaspectratio]{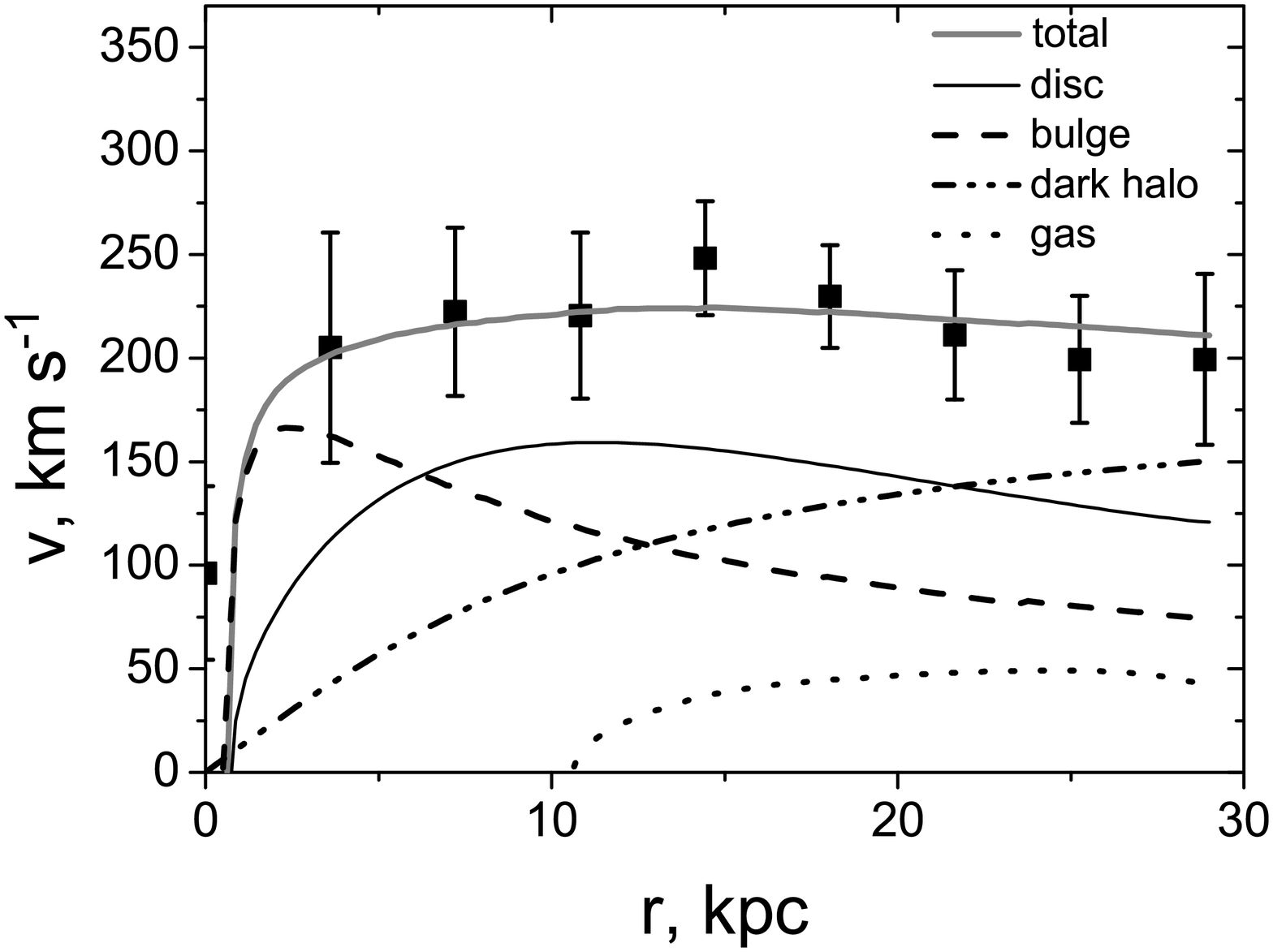}
\includegraphics[width=0.4\textwidth,keepaspectratio]{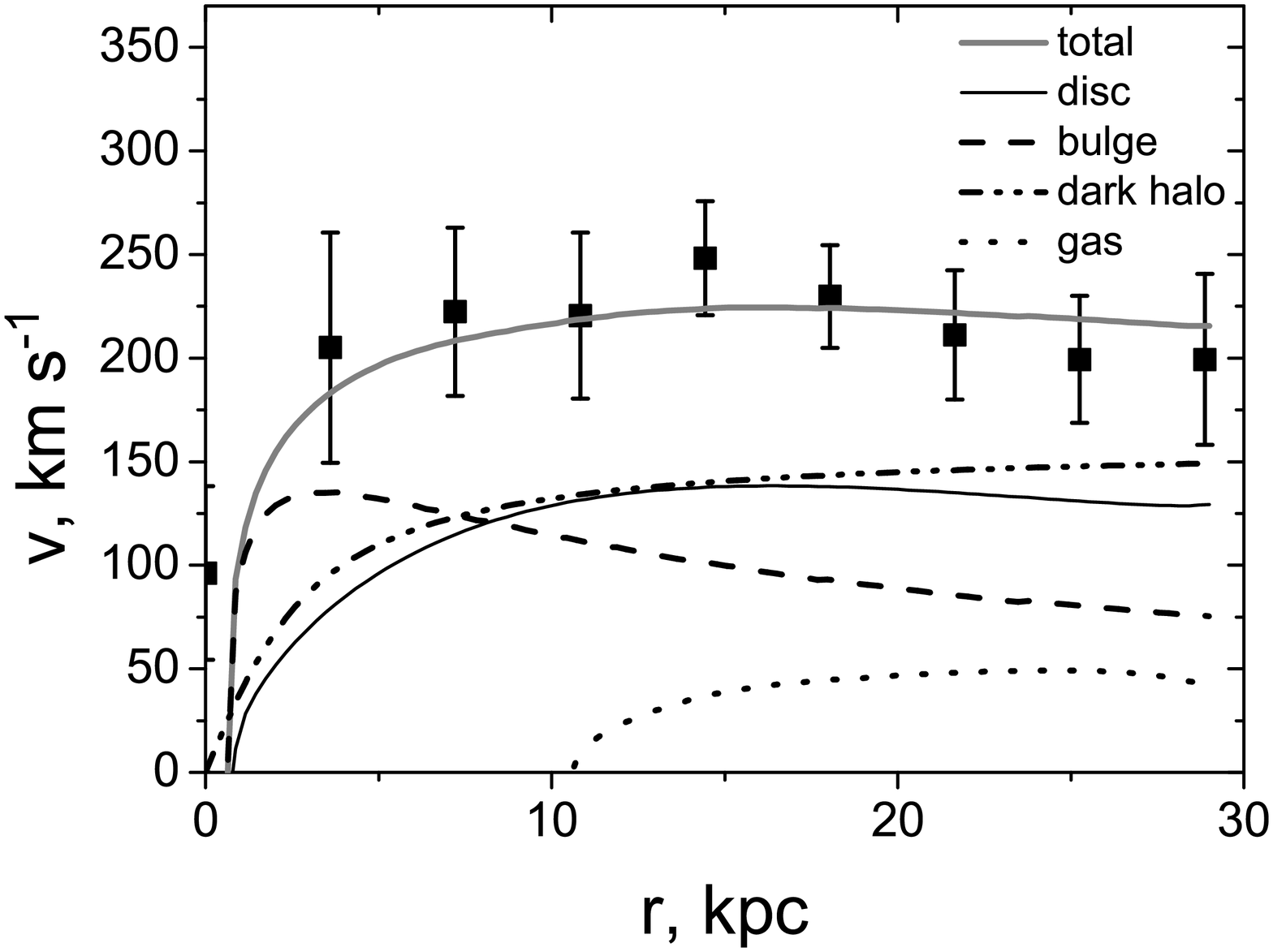}

\caption{The results of mass modeling of UGC11919 assuming $i=13 \degr$ for the disc and bulge mass-to-light ratios obtained from the color indices in Paper I (varying with radius) (top panel) and from the SED and spectral fitting in current article (constant with radius) (bottom panel) . } 
\label{rcmod}
\end{figure}

We performed the decomposition of the rotation curve obtained from the reprocessing of the H{\sc i}  data into Sersic bulge, exponential stellar disc, gaseous disc and pseudo-isothermal dark
  halo. The density profile of the H{\sc i}   disc is taken
from the observations and scaled by 1.3 to include the helium. 
 The radial profiles of surface density of stellar disc and bulge are proportional to the distribution of their surface brightness (from Paper I). Two models were considered. In Model 1 the mass-to-light ratios of disc and bulge are calculated using the $M/L$-color model relations for scaled Salpeter IMF of \cite{bdj}. They are  varying with radius according to the change of the color index $(B-V)_0$ of bulge and disc obtained in Paper I. In the Model 2 we use the constant mass-to-light ratios of stellar components derived from the spectra and SED modeling in Sect.\ref{ss}. As long as the surface densities of visible components are fixed the parameters of the dark matter halo (radial scale and asymptotic velocity) were scaled to
achieve the minimal deviation of the resulting model
rotation curve from the observed curve. Both best fit models of the rotation curve are shown in Fig. \ref{rcmod}. As could be seen from Fig. \ref{rcmod} both models give a good fit to the obtained rotation curve. In Table \ref{tablercmod} we give the masses of the components and the total mass ($\times 10^{10} M_{\odot}$) within the radius $r$ for the two models. The dark halo contains more than a half of the total mass within the last measured radius of the rotation curve for both models.
\begin{table*}
\small \caption{The results of mass modeling. 
\label{tablercmod}}
  \begin{center}
    \begin{tabular}{c c c c c c c }
    \hline 
&$r$, kpc & $M_{ disc}$, $\times 10^{10} M_{\odot}$    & $M_{ halo}$, $\times 10^{10} M_{\odot}$  & $M_{ bulge}$, $\times 10^{10} M_{\odot}$  & $M_{ gas}$, $\times 10^{10} M_{\odot}$  &$M_{ tot}$, $\times 10^{10} M_{\odot}$ \\

\hline
Model 1&29&7.8&15&3.7&0.9&27.4\\ 

Model 2&29&7.6&15&3.8&0.9&27.3\\
\hline
\end{tabular}
  \end{center}
\end{table*}

\section{Conclusions}

We performed the long-slit observations of spiral galaxy UGC11919 which was suspected in
the previous study (see Paper I) as a galaxy possessing a peculiarly low dynamical disc and
bulge mass-to-light ratios. The main results are given below.
\begin{itemize}
 \item We estimated the radial profiles of line-of-sight
velocities and the  velocity dispersion for stars and gas. Spectral slices of UGC11919 allow to
propose that the galaxy possesses a nuclear disc, where the stellar velocity dispersion is lower than in the bulge. 

 \item The analysis of long-slit data
together with SED with scaled Salpeter and Kroupa IMFs  confirms the previous conclusion that the mass-to-light ratios of stellar population  
are several times higher than it follows from the dynamic
modeling of the rotation curve if the disc inclination angle corresponds to the photometrically defined values $i\geq 30 \degr$. However the re-processing of H{\sc i} data showed that the lower inclination i = $13 \degr$ cannot be excluded from dynamical data, which may eliminate the contradiction between the dynamical and photometrical estimations of disc  mass. In addition, the stellar velocity dispersion at the radial scalelength $r_d$ better agrees with the ''heavy'' disc (i = $13 \degr$) -- the ''light'' disc corresponds to the unrealistically large thickness of the disc $Z_0/r_d=0.7$. The lower inclination also is in a good agreement with the position of the galaxy at the Tully-Fisher diagrams. It allows to conclude  that the disc inclination
of UGC11919 is significantly lower than it follows from the photometry
which may  evidence the non-round  shape of the disc even after the correction for two-armed spiral structure.

\item Applying the sophisticated models of
stellar population with ``normal'' (Salpeter and Kroupa) stellar IMFs to the spectra and the broad-band SED we
obtained the mean age (4.2, 2.6 and 2.3 Gyr) and 
metallicity [Z/H]  (-0.4, -0.5  and -0.3 dex) for the bulge, disc and nuclear disc of UGC11919 correspondingly. It reveals the long lasting star formation in all three components.

\item We decomposed the rotation curve and estimated the masses of bulge, disc and halo assuming photometry based $M/L$ of stellar components.

\item We found that the main disc of the galaxy is dynamically overheated. In principle, this could be the result of gravitation interaction with the companions, some of which are observed only in H{\sc i}  (see Paper I).

\end{itemize}
{\bf Acknowledgments} We thank the anonymous referee for the important remarks which allowed us to improve the paper.
The observations at the 6-meter BTA telescope were carried out
with the financial support of the Ministry of Education and Science of the Russian Federation
(agreement No. 14.619.21.0004, project ID RFMEFI61914X0004). The analysis and interpretation of the long-slit observational data was supported by the Russian Science
Foundation grant  no.  14-22-00041. The reprocessing of the 21 cm data is made with the support of the RFBR, research project No. 15-32-21062 a.

We are grateful to Gyula J\'ozsa for providing the TiRiFiC
software and for the fruitful discussion.
We acknowledge the usage of the HyperLeda database (http://leda.univ-lyon1.fr).
The funding for the SDSS has been provided by the Alfred P. Sloan Foundation, the Participating Institutions, the
National Science Foundation, the United States Department
of Energy, the National Aeronautics and Space Administration, the Japanese Monbukagakusho, the Max Planck Society and the Higher Education Funding Council for England.
\bibliographystyle{mn2e}
\bibliography{saburova}

\begin{thebibliography}{}

\bibitem[\protect\citeauthoryear{{Afanasiev} \& {Moiseev}}{{Afanasiev} \&
  {Moiseev}}{2005}]{AfanasievMoiseev2005}
{Afanasiev} V.~L.,  {Moiseev} A.~V.,  2005, Astronomy Letters, 31, 194

\bibitem[\protect\citeauthoryear{{Bastian}, {Covey} \& {Meyer}}{{Bastian}
  et~al.}{2010}]{Bastian}
{Bastian} N.,  {Covey} K.~R.,    {Meyer} M.~R.,  2010, \araa, 48, 339

\bibitem[\protect\citeauthoryear{{Bell} \& {de Jong}}{{Bell} \& {de
  Jong}}{2001}]{bdj}
{Bell} E.~F.,  {de Jong} R.~S.,  2001, \apj, 550, 212

\bibitem[\protect\citeauthoryear{{Binney} \& {Tremaine}}{{Binney} \&
  {Tremaine}}{2008}]{BinneyTremaine}
{Binney} J.,  {Tremaine} S.,  2008, {Galactic Dynamics: Second Edition}.
Princeton University Press

\bibitem[\protect\citeauthoryear{{Chilingarian} \& {Katkov}}{{Chilingarian} \&
  {Katkov}}{2012}]{Chilingarian2012}
{Chilingarian} I.~V.,  {Katkov} I.~Y.,  2012, in {Tuffs} R.~J.,  {Popescu}
  C.~C.,  eds, IAU Symposium Vol.~284 of IAU Symposium, {NBursts+phot:
  parametric recovery of galaxy star formation histories from the simultaneous
  fitting of spectra and broad-band spectral energy distributions}.
pp 26--28

\bibitem[\protect\citeauthoryear{{Chilingarian}, {Prugniel}, {Sil'Chenko} \&
  {Afanasiev}}{{Chilingarian} et~al.}{2007}]{Chilingarian2007}
{Chilingarian} I.~V.,  {Prugniel} P.,  {Sil'Chenko} O.~K.,    {Afanasiev}
  V.~L.,  2007, \mnras, 376, 1033

\bibitem[\protect\citeauthoryear{{Gilmore}}{{Gilmore}}{2001}]{Gilmore}
{Gilmore} G.,  2001, in {Tacconi} L.,  {Lutz} D.,  eds, Starburst Galaxies:
  Near and Far {Evidence Supporting the Universality of the IMF}.
p.~34

\bibitem[\protect\citeauthoryear{{Gunawardhana}, {Hopkins}, {Sharp}, {Brough},
  {Taylor}, {Bland-Hawthorn} \& {Maraston}}{{Gunawardhana}
  et~al.}{2011}]{Gunawardhana}
{Gunawardhana} M.~L.~P.,  {Hopkins} A.~M.,  {Sharp} R.~G.,  {Brough} S.,
  {Taylor} E.,  {Bland-Hawthorn} J.,    {Maraston} C. e.~a.,  2011, \mnras,
  415, 1647

\bibitem[\protect\citeauthoryear{{J{\'o}zsa}, {Kenn}, {Klein} \&
  {Oosterloo}}{{J{\'o}zsa} et~al.}{2007}]{J2007}
{J{\'o}zsa} G.~I.~G.,  {Kenn} F.,  {Klein} U.,    {Oosterloo} T.~A.,  2007,
  \aap, 468, 731

\bibitem[\protect\citeauthoryear{{Kasparova}, {Saburova}, {Katkov},
  {Chilingarian} \& {Bizyaev}}{{Kasparova} et~al.}{2014}]{Kasparova2014}
{Kasparova} A.~V.,  {Saburova} A.~S.,  {Katkov} I.~Y.,  {Chilingarian} I.~V.,
   {Bizyaev} D.~V.,  2014, \mnras, 437, 3072

\bibitem[\protect\citeauthoryear{{Katkov} \& {Chilingarian}}{{Katkov} \&
  {Chilingarian}}{2011}]{Katkov2011}
{Katkov} I.~Y.,  {Chilingarian} I.~V.,  2011, in {Evans} I.~N.,  {Accomazzi}
  A.,  {Mink} D.~J.,   {Rots} A.~H.,  eds, Astronomical Data Analysis Software
  and Systems XX Vol.~442 of Astronomical Society of the Pacific Conference
  Series, {A New Sky Subtraction Technique for Low Surface Brightness Data}.
p.~143

\bibitem[\protect\citeauthoryear{{Khoperskov}, {Zasov} \&
  {Tyurina}}{{Khoperskov} et~al.}{2003}]{Khoperskov2003}
{Khoperskov} A.~V.,  {Zasov} A.~V.,    {Tyurina} N.~V.,  2003, Astronomy
  Reports, 47, 357

\bibitem[\protect\citeauthoryear{{Kroupa}}{{Kroupa}}{2001}]{imf2}
{Kroupa} E.,  2001, \mnras, 322, 231

\bibitem[\protect\citeauthoryear{{Kroupa}}{{Kroupa}}{2002}]{Kroupa2002}
{Kroupa} P.,  2002, Science, 295, 82

\bibitem[\protect\citeauthoryear{{Le Borgne}, {Rocca-Volmerange}, {Prugniel},
  {Lan{\c c}on}, {Fioc} \& {Soubiran}}{{Le Borgne}
  et~al.}{2004}]{LeBorgneetal2004}
{Le Borgne} D.,  {Rocca-Volmerange} B.,  {Prugniel} P.,  {Lan{\c c}on} A.,
  {Fioc} M.,    {Soubiran} C.,  2004, \aap, 425, 881

\bibitem[\protect\citeauthoryear{{McGaugh}}{{McGaugh}}{2005}]{McGaugh2005}
{McGaugh} S.~S.,  2005, \apj, 632, 859

\bibitem[\protect\citeauthoryear{{McGaugh} \& {Schombert}}{{McGaugh} \&
  {Schombert}}{2015}]{McGaugh2015}
{McGaugh} S.~S.,  {Schombert} J.~M.,  2015, \apj, 802, 18

\bibitem[\protect\citeauthoryear{{Prugniel}, {Soubiran}, {Koleva} \& {Le
  Borgne}}{{Prugniel} et~al.}{2007}]{Prugniel2007}
{Prugniel} P.,  {Soubiran} C.,  {Koleva} M.,    {Le Borgne} D.,  2007, ArXiv
  Astrophysics e-prints

\bibitem[\protect\citeauthoryear{{Romeo} \& {Wiegert}}{{Romeo} \&
  {Wiegert}}{2011}]{Romeo2011}
{Romeo} A.~B.,  {Wiegert} J.,  2011, \mnras, 416, 1191

\bibitem[\protect\citeauthoryear{{Saburova}, {J{\'o}zsa}, {Zasov} \&
  {Bizyaev}}{{Saburova} et~al.}{2013}]{saburovaetal2013}
{Saburova} A.~S.,  {J{\'o}zsa} G.~I.~G.,  {Zasov} A.~V.,    {Bizyaev} D.~V.,
  2013, \aap, 554, A128

\bibitem[\protect\citeauthoryear{{Saburova}, {Shaldenkova} \&
  {Zasov}}{{Saburova} et~al.}{2009}]{Saburova2009}
{Saburova} A.~S.,  {Shaldenkova} E.~S.,    {Zasov} A.~V.,  2009, Astronomy
  Reports, 53, 801

\bibitem[\protect\citeauthoryear{{Saburova} \& {Zasov}}{{Saburova} \&
  {Zasov}}{2012}]{Saburova2012}
{Saburova} A.~S.,  {Zasov} A.~V.,  2012, Astronomy Letters, 38, 139

\bibitem[\protect\citeauthoryear{{Salpeter}}{{Salpeter}}{1955}]{imf}
{Salpeter} E.,  1955, \apj, 121, 161

\bibitem[\protect\citeauthoryear{{Shapiro}, {Gerssen} \& {van der
  Marel}}{{Shapiro} et~al.}{2003}]{Shapiro2003}
{Shapiro} K.~L.,  {Gerssen} J.,    {van der Marel} R.~P.,  2003, \aj, 126, 2707

\bibitem[\protect\citeauthoryear{{van der Marel} \& {Franx}}{{van der Marel} \&
  {Franx}}{1993}]{vanderMarel}
{van der Marel} R.~P.,  {Franx} M.,  1993, \apj, 407, 525

\bibitem[\protect\citeauthoryear{{Zasov}, {Khoperskov} \& {Saburova}}{{Zasov}
  et~al.}{2011}]{Zasov2011}
{Zasov} A.~V.,  {Khoperskov} A.~V.,    {Saburova} A.~S.,  2011, Astronomy
  Letters, 37, 374

\bibitem[\protect\citeauthoryear{{Zasov}, {Khoperskov} \& {Tyurina}}{{Zasov}
  et~al.}{2004}]{Zasov2004}
{Zasov} A.~V.,  {Khoperskov} A.~V.,    {Tyurina} N.~V.,  2004, Astronomy
  Letters, 30, 593

\end{thebibliography}
\end{document}